\documentclass{aastex631}
\usepackage{epstopdf}
\usepackage{bm}

\newcommand\aastex{AAS\TeX}

\received{xx}
\revised{xx}
\accepted{xx}
\submitjournal{ApJ}

\shorttitle{\aastex\ sample article}
\shortauthors{Cai}

\begin{document}

\title{Large-scale Vortices in Rapidly Rotating Rayleigh-B\'enard Convection at Small Prandtl number}

\correspondingauthor{Tao Cai}
\email{tcai@must.edu.mo}

\author[0000-0003-3431-8570]{Tao Cai}
\affil{State Key Laboratory of Lunar and Planetary Sciences, Macau University of Science and Technology, Macau, P.R.China}



\begin{abstract}
One prominent feature in the atmospheres of Jupiter and Saturn is the appearance of large-scale vortices. However, the sustaining mechanism of these large-scale vortices remains unclear. One possible mechanism is that these large-scale vortices are driven by rotating convection. Here we present numerical simulation results on rapidly rotating Rayleigh-B\'enard convection at a small Prandtl number $Pr=0.1$ (close to the turbulent Prandtl numbers of Jupiter and Saturn). We have identified four flow regimes in our simulation: multiple small vortices, coexisted large-scale cyclone and anticyclone, large-scale cyclone, and turbulence. The formation of large-scale vortices requires two conditions to be satisfied: the vertical Reynolds number is large ($Re_{z}\ge 400$), and the Rossby number is small ($Ro\leq 0.4$). Large-scale cyclone first appears when $Ro$ decreases to be smaller than 0.4. When $Ro$ further decreases to be smaller than 0.1, coexisted large-scale anticyclone emerges. We have studied the heat transfer in rapidly rotating convection. The result reveals that the heat transfer is more efficient in the anticyclonic region than in the cyclonic region. Besides, we find that 2D effect increases and 3D effect decreases in transporting convective flux as rotation rate increases. We find that aspect ratio has an effect on the critical Rossby number for the emergence of large-scale vortices. Our results provide helpful insights on understanding the dynamics of large-scale vortices in gas giants.
\end{abstract}




\section{Introduction} \label{sec:intro}
Rotation plays an important role on stellar and planetary turbulent convection. It has been observed that long-lived large-scale vortices (LSVs) are formed in rapidly rotating giant planets in our solar system. At the atmosphere of the Jupiter, for example, large-scale Great Red Spot has been observed for more than 180 years. Recently, large-scale collective cyclones have been observed in polar regions of Jupiter by the Juno spacecraft \citep{adriani2018clusters,tabataba2020long,adriani2020two}. Despite collective cyclones, dipole configurations of large-scale cyclone and anticyclone have also been observed in the polar regions of Jupiter \citep{adriani2020two}. For Saturn, \citet{godfrey1988hexagonal} reported that a large-scale hexagonal cyclone has been observed in the north pole by the Cassini spacecraft. Later, \citet{vasavada2006cassini} reported that a large-scale cyclone also exists in the south pole of Saturn. Apart from polar vortices, various cyclones and anticyclones were observed in other regions of Jupiter \citep{li2004life} and Saturn \citep{sayanagi2013dynamics}. However, the driving mechanism of these vortices remains unclear. There are two scenarios for the explanations: one is based on shallow models \citep{zhang2014atmospheric,o2016weak,brueshaber2019dynamical} which explain that these vortices can be formed by merging of random storms generated by moist convection; and the other is based on deep models \citep{chan2013numerical,yadav2020deep,cai2021deep} which explain that these vortices are generated from rotating turbulent convection powered by internal heat. Progresses have been made in both types of models recently. For example, based on a shallow model, \citet{brueshaber2019dynamical} searched the parameter space and found that Burger number (square of the ratio of the Rossby deformation radius to planetary radius) is a key factor to determine the pattern of polar vortices. On the other hand, based on a deep model, \citep{cai2021deep} has successfully replicated the pentagonal and hexagonal patterns of circumpolar vortices observed on the south pole of Jupiter. It indicates that rapidly rotating turbulent convection might be an important mechanism on generating vortices in gas giant planets. Apart from gas giant planets, LSVs are possibly also prominent features in rapidly rotating stars. LSVs-like star spots in cool stars, have been observed in stellar atmospheres in Doppler imaging maps \citep{hackman2019starspot,willamo2019long}. The star spots observed in cool stars could be associated with the LSVs driven by rotation \citep{kapyla2011starspots}.
In this paper, we will mainly focus on the LSVs formed in rapidly rotating turbulent convection.

Early numerical simulation \citep{chan2007rotating} on compressible convection in Cartesian geometry has discovered that LSVs could be generated when the rotation is fast. This phenomena has been confirmed in subsequent works on rapidly rotating compressible convection \citep{mantere2011dependence,kapyla2011starspots,chan2013numerical,cai2016semi,cai2021deep}.  For simulations on incompressible flow, LSVs were first reported in \citet{julien2012statistical}, where a set of reduced rapidly rotating Rayleigh-B\'enard equations were solved. The appearance of LSVs was also observed in direct numerical simulations on rapidly rotating Rayleigh-B\'enard convection (RRBC) \citep{favier2014inverse,guervilly2014large,guervilly2015generation,kunnen2016transition,guervilly2017jets,novi2019rapidly}. In these simulations, large-scale cyclone appears with associated large-scale circulation of small anticyclonic vortices. The flow patterns of coexisted large-scale cyclone and anticyclone in RRBC were reported when the Rayleigh number achieves $2 \times 10^{11}$ at $Pr=1$ \citep{stellmach2014approaching} with stress-free boundary condition and $1.5 \times 10^{11}$ at $Pr=5.2$ with no-slip boundary condition \citep{guzman2020competition}. In spherical geometry, large-scale vortices were also found recently in anelastic convection \citep{yadav2020deep} and incompressible convection \citep{lin2021large}.

So far, most of the studies on RRBC were focused on the moderate Prandtl number region $Pr \sim O(1)$. However, the Prandtl numbers in stars or gas giant planets are usually smaller than one \citep{kupka2017modelling,schubert2011planetary}. Linear instability analysis shows that the fluid at small $Pr$ behaves different from that at large $Pr$ \citep{chandrasekhar2013hydrodynamic,zhang1987onset}. For example, the onset of instability first occurs as oscillatory convection when $Pr<0.67$ in the fluid on an infinite planes \citep{chandrasekhar2013hydrodynamic}. The critical Rayleigh number $Ra_{c}$ for the onset of convection is much lower at low $Pr$ ($Ra_{c}\sim 17.4 Pr^{4/3}(1+Pr)^{-1/3} E^{-4/3}$) than at high $Pr$ ($Ra_{c}\sim 8.7 E^{-4/3}$) \citep{chandrasekhar2013hydrodynamic}, where $E$ is the Ekman number. In addition, experiments on liquid metal gallium ($Pr=0.025$) with and without rotation have shown that the convective behavior of low $Pr$ is substantially different from that of moderate $Pr$ \citep{king2013turbulent}. At moderate $Pr$, four flow regimes are identified: cells, convective Taylor column, plumes, and geostrophic turbulence \citep{julien2012statistical}. The trend toward lower $Pr$ indicates that geostrophic turbulent regime is approaching at low $Pr$ \citep{aurnou2015rotating}. Compressible simulations on rapidly rotating convection \citep{kapyla2011starspots,chan2013numerical,cai2016semi} showed that at low $Pr$ the flow favours in the formation of LSVs. Hence we expect that the flow regimes would be different from those identified in the simulations at moderate $Pr$. Previous simulations on Rayleigh-B\'enard convection at $Pr=0.1$ with no-slip boundary condition found large-scale cyclones when the rotation is fast \citep{guzman2020competition}. In this paper, we will explore the flow regimes of RRBC at low $Pr$ with stress-free boundary condition through numerical simulations. We find that the flow pattern of coexisted large-scale cyclone and anticyclone appears in RBBC. We will also discuss the energy and heat transfer among different regimes.

\section{The model}
For the rotating Rayleigh-B\'enard convection in a Cartesian box, the nondimensional hydrodynamic equations describing mass, momentum, energy conservations could be written as
\begin{eqnarray}
\bm{\nabla} \bm{\cdot} \bm{u} &=&0~,\label{eq1}\\
\partial_{t}\bm{u}  &=& -\bm{u} \bm{\cdot} \bm{\nabla} \bm{u}-\nabla p +\Theta \hat{\bm{z}} + \sqrt{\frac{Pr}{Ra}} \nabla^2 \bm{u}- Ro^{-1}\hat{\bm{z}} \bm{\times} \bm{u} ~,\label{eq2}\\
\partial_{t} \Theta&=&-\bm{u} \bm{\cdot} \nabla\Theta-\bm{u}\bm{\cdot} \bm{\nabla}T_{s}+\sqrt{\frac{1}{RaPr}}\nabla^2 \Theta~\label{eq3},
\end{eqnarray}
where $\bm{u}$ is the velocity, $p$ is the reduced pressure, $\hat{\bm{z}}$ is the unit vector in the vertical direction, $\Theta$ is the superadiabatic temperature, $T_{s}$ is the static reference state of the temperature, $\nu$ is the kinematic viscosity, $\kappa$ is the thermometric conductivity, $g$ is the gravitational acceleration, $\alpha$ is the coefficient of volume expansion, $H$ is the height of the box, $\delta T$ is temperature difference between the bottom and top of the box, $\Omega$ is the angular velocity, $Pr=\nu/\kappa$ is the Prandtl number, $Ra=g\alpha \delta T H^3/(\nu \kappa)$ is the Rayleigh number, $E=\nu/(2\Omega H^2)$ is the Ekman number, $Ro=Re E$ is the convective Rossby number. We also define the Reynolds number $Re=\sqrt{RaPr^{-1}}$, and the P\'{e}clet number $Pe=RePr$. In the above equations, the normalizing factors for length, time, velocity, pressure, and superadiabatic temperature are $H$, $Re^{-1}H^2/\nu$, $Re \nu/H$, $Re^2 \rho \nu^2/H^2$, and $\delta T$, respectively.

To solve the incompressible equations, the velocity is decomposed into
\begin{eqnarray}
\boldsymbol{u}=\boldsymbol{\nabla}\boldsymbol{\times}\boldsymbol{\nabla}\boldsymbol{\times}(\Phi \hat{\boldsymbol{z}})+\boldsymbol{\nabla}\boldsymbol{\times}(\Psi \hat{\boldsymbol{z}})~,
\end{eqnarray}
where $\Phi$ and $\Psi$ are the poloidal and toroidal potentials.
Applying the operators $\widehat{\bm{z}} \bm{\cdot} \bm{\nabla} \times$ and $\widehat{\bm{z}} \bm{\cdot} \bm{\nabla} \times \bm{\nabla} \times$ to the momentum equation, we have
\begin{eqnarray}
\Delta_{h}[\partial_{t}\Psi -Re^{-1}\Delta\Psi-Ro^{-1} \partial_{z}\Phi]&=& \bm{\nabla}_{h}\bm{\times} \bm{G}_{h}~,\label{eq4}\\
\Delta_{h}[\partial_{t}\Delta\Phi-  Re^{-1}\Delta\Delta\Phi+Ro^{-1}\partial_{z}\Psi+\Theta]&=& -\bm{\nabla}_{h} \bm{\cdot} \partial_{z}\bm{G}_{h}+\Delta_{h}G_{z}~,\label{eq5}\\
\partial_{t}\Theta-Pe^{-1}\Delta\Theta-\frac{dT_{s}}{dz}\Delta_{h}\Phi&=&-\bm{\nabla}\bm{\cdot}(\Theta\bm{u})~,\label{eq6}
\end{eqnarray}
where the subscript $h$ denotes the horizontal component, $\Delta$ is the Laplacian operator, $\boldsymbol{\omega}$ is the vorticity, and $\boldsymbol{G}=\boldsymbol{\omega}\times\boldsymbol{u}$ is a nonlinear term. Equations (\ref{eq4}-\ref{eq6}) are solved numerically using a mixed finite-difference pseudo-spectral method. Fourier transforms are applied in the horizontal directions, and a second-order finite difference method is adopted in the vertical direction. A 2/3 dealiasing rule was used for the pseudo-spectral scheme in the horizontal direction. All the linear terms on the l.h.s. of equations (\ref{eq4}-\ref{eq6}) are integrated by a second-order semi-implicit scheme. The quadratic nonlinear terms on the r.h.s. of equations (\ref{eq4}-\ref{eq6}) are integrated by a third-order explicit Adams-Bashforth scheme. The numerical method is similar to the one developed in \citet{cai2016semi}, where a compressible flow is considered instead. The upper and lower boundary conditions are taken to be thermally conducting, impenetrable and stress-free. The lateral boundary conditions on both sides are periodic. The grid points on the horizontal directions are uniformly distributed. On the vertical direction, Chebyshev-Gauss-Lobatto grid points are used to resolve the boundary layers. The lateral-to-height aspect ratios are set to be unity ($\Gamma=1$). In this paper, we choose a small fixed Prandtl number $Pr=0.1$. Three groups of simulations with different Rayleigh numbers $Ra=10^6$, $10^7$, and $10^8$ are performed. In each group, the reduced Rayleigh numbers $\widetilde{Ra}=Ra E^{4/3}$ are varied from 1 to 1000 for comparisons. $\widetilde{Ra}$ is related to the supercritical Rayleigh number $Ra_{c}\sim O(E^{-4/3})$ \citep{chandrasekhar2013hydrodynamic,julien2012statistical}. For a rotating B\'enard flow with $Pr>0.67$ in a layer, the onset of instability first occurs as stationary convection when $Ra>Ra_{c1}=8.7E^{-4/3}$ \citep{chandrasekhar2013hydrodynamic}. For $Pr<0.67$, the onset of instability first occurs as oscillatory convection when $Ra>Ra_{c2}=17.4 Pr^{4/3}(1+Pr)^{-1/3} E^{-4/3}$ \citep{chandrasekhar2013hydrodynamic}. In this paper, we mainly focus on the small Prandtl number cases at $Pr=0.1$, which yields $Ra_{c2}=0.78E^{-4/3}$. Since $Ra_{c2}$ is smaller than $Ra_{c1}$ by an order of magnitude, we can explore lower Ekman number regimes with moderate Rayleigh numbers in our simulation. The grid resolutions are $N_{x}\times N_{y}\times N_{z}=512\times 512 \times 257$ for the cases with $Ra=10^8$ (group C), and $N_{x}\times N_{y}\times N_{z}=256\times 256 \times 257$ for the cases with $Ra=10^6$ (group A) and $Ra=10^7$ (group B). In a low Prandtl flow, the thermal boundary layer is estimated to has a thickness of $\delta_{T}\approx (Ra Pr)^{-1/4}$ \citep{horanyi1999turbulent,king2013turbulent}. With a Chebyshev grid distribution in the vertical direction, there are about 40, 30, and 22 grid points within the thermal boundary layers for the cases in groups A, B, and C, respectively. The detailed parameters of simulation cases are shown in Table~\ref{table:tab1}. The critical wavelength $\ell_{c}$ at the onset of convection is also reported for reference. The critical wavelength $\ell_{c}\approx 2^{7/6}\pi^{2/3}[Pr/(1+Pr)]^{-1/3} E^{1/3}$ for a rapidly rotating flow at $Pr<0.67$, and $\ell_{c}\approx 2^{7/6}\pi^{2/3} E^{1/3}$ for a rapidly rotating flow at $Pr>0.67$ \citep{chandrasekhar2013hydrodynamic}. In each case, we run the simulation for a long time till the system reaches a statistically thermal relaxation state. In practice, we require that the variation of the averaged Nusselt number (the average is taken temporally and horizontally) is smaller than one percent. The time for reaching statistically relaxation state is different for different cases. For some typical cases, such as C3 and C5, we have run for a period of about 30000 units of time (about one unit of viscous dissipative timescale).

\startlongtable
\begin{deluxetable*}{ccccccccccccccccccc}
\tablecaption{Parameters of simulation cases at $Pr=0.1$\label{table:tab1}}
\tablehead{
 Case & $\Gamma$ & $\ell_{c}$  & $u''$ & $u_{z}''$ & $u_{h}''$ & $Ra$ & $Re_{z}$ & $Ro$ & $E$ & $\widetilde{Ra}$ & $Nu$ & Regime
}
\startdata
A1 & 1 & 0.339 & 0.0342 & 0.0183 & 0.0296 & $10^6$  & 57.84   & 0.1    &  $3.162\times 10^{-5}$ & 1         & 1.037  & I   \\
A2 & 1 & 0.403 & 0.1060 & 0.0379 & 0.0973 & $10^6$  & 119.82  & 0.1682  & $5.318\times 10^{-5}$ & 2         & 1.171  & I   \\
A3 & 1 & 0.479 & 0.2251 & 0.0845 & 0.2067 & $10^6$  & 267.36  & 0.2828  & $8.944\times 10^{-5}$ & 4         & 2.024  & I   \\
A4 & 1 & 0.570 & 0.2439 & 0.1389 & 0.1960 & $10^6$  & 439.40  & 0.4757  & $1.504\times 10^{-4}$ & 8         & 4.013  & IV   \\
A5 & 1 & 0.602 & 0.2496 & 0.1472 & 0.1968 & $10^6$  & 465.50  & 0.5623  & $1.778\times 10^{-4}$ & 10        & 4.548  & IV   \\
A6 & 1 & 0.716 & 0.2751 & 0.1729 & 0.2087 & $10^6$  & 546.65  & 0.9457  & $2.991\times 10^{-4}$ & 20        & 6.073  & IV   \\
A7 & 1 & 0.793 & 0.2930 & 0.1877 & 0.2191 & $10^6$  & 593.69  & 1.2819  & $4.054\times 10^{-4}$ & 30        & 6.873  & IV   \\
A8 & 1 & 0.852 & 0.3107 & 0.2046 & 0.2273 & $10^6$  & 646.94  & 1.5905  & $5.030\times 10^{-4}$ & 40        & 7.535  & IV   \\
A9 & 1 & 0.901 & 0.3305 & 0.2221 & 0.2371 & $10^6$  & 702.40  & 1.8803  & $5.946\times 10^{-4}$ & 50        & 8.221  & IV   \\
A10 & 1 & 0.943 & 0.3365 & 0.2278 & 0.2398 & $10^6$  & 720.22  & 2.1558  & $6.817\times 10^{-4}$ & 60        & 8.488  & IV   \\
A11 & 1 & 0.980 & 0.3517 & 0.2417 & 0.2466 & $10^6$  & 746.20  & 2.4200  & $7.653\times 10^{-4}$ & 70        & 9.018  & IV   \\
A12 & 1 & 1.013 & 0.3585 & 0.2487 & 0.2487 & $10^6$  & 786.34  & 2.6750  & $8.459\times 10^{-4}$ & 80        & 9.297  & IV   \\
A13 & 1 & 1.043 & 0.3660 & 0.2556 & 0.2519 & $10^6$  & 808.34  & 2.9220  & $9.240\times 10^{-4}$ & 90        & 9.478  & IV   \\
A14 & 1 & 1.071 & 0.3731 & 0.2616 & 0.2554 & $10^6$  & 827.14  & 3.1623  & $1.000\times 10^{-3}$ & 100       & 9.912  & IV   \\
A15 & 1 & 1.274 & 0.4111 & 0.2968 & 0.2699 & $10^6$  & 759.90  & 5.3183  & $1.682\times 10^{-3}$ & 200       & 11.249 & IV   \\
A16 & 1 & 1.602 & 0.4281 & 0.3124 & 0.2759 & $10^6$  & 987.96  & 10.5737 & $3.344\times 10^{-3}$ & 500       & 12.088 & IV   \\
A17 & 1 & 1.905 & 0.4389 & 0.3213 & 0.2808 & $10^6$  & 1015.96  & 17.7828 & $5.623\times 10^{-3}$ & 1000      & 12.514 & IV   \\
A18 & 1 & 2.221 & 0.4472 & 0.3297 & 0.2827 & $10^6$  & 1042.50  & $\infty$& $\infty$              & $\infty$  & 12.780 & IV   \\
\hline
B1 & 1 & 0.191 & 0.0227 & 0.0106 & 0.0192 & $10^7$  & 106.38  & 0.0562  & $5.623\times 10^{-6}$ & 1         & 1.034  & I   \\
B2 & 1 & 0.227 & 0.0912 & 0.0228 & 0.0877 & $10^7$  & 227.76  & 0.0946  & $9.457\times 10^{-6}$ & 2         & 1.176  & I   \\
B3 & 1 & 0.269 & 0.2136 & 0.0550 & 0.2057 & $10^7$  & 550.22  & 0.1591  & $1.591\times 10^{-5}$ & 4         & 2.126  & III   \\
B4 & 1 & 0.320 & 0.3099 & 0.0982 & 0.2931 & $10^7$  & 982.12  & 0.2675  & $2.675\times 10^{-5}$ & 8         & 5.232  & III   \\
B5 & 1 & 0.339 & 0.2843 & 0.1065 & 0.2626 & $10^7$  & 1064.96  & 0.3162  & $3.162\times 10^{-5}$ & 10       & 6.299  & III   \\
B6 & 1 & 0.398 & 0.2225 & 0.1329 & 0.1751 & $10^7$  & 1329.12 & 0.5318  & $5.138\times 10^{-5}$ & 20        & 9.731  & IV   \\
B7 & 1 & 0.446 & 0.2372 & 0.1447 & 0.1844 & $10^7$  & 1446.90 & 0.7208  & $7.208\times 10^{-5}$ & 30        & 11.783 & IV   \\
B8 & 1 & 0.479 & 0.2462 & 0.1534 & 0.1889 & $10^7$  & 1533.87 & 0.8944  & $8.944\times 10^{-5}$ & 40        & 12.948 & IV   \\
B9 & 1 & 0.506 & 0.2573 & 0.1620 & 0.1959 & $10^7$  & 1619.53 & 1.0574  & $1.057\times 10^{-4}$ & 50        & 14.024 & IV   \\
B10 & 1 & 0.530 & 0.2620 & 0.1672 & 0.1975 & $10^7$  & 1672.35 & 1.2123  & $1.212\times 10^{-4}$ & 60        & 14.580 & IV   \\
B11 & 1 & 0.551 & 0.2716 & 0.1755 & 0.2027 & $10^7$  & 1755.21 & 1.3609  & $1.361\times 10^{-4}$ & 70        & 15.393 & IV   \\
B12 & 1 & 0.570 & 0.2796 & 0.1819 & 0.2075 & $10^7$  & 1818.56 & 1.5042  & $1.504\times 10^{-4}$ & 80        & 15.969 & IV   \\
B13 & 1 & 0.587 & 0.2819 & 0.1845 & 0.2082 & $10^7$  & 1845.33 & 1.6432  & $1.643\times 10^{-4}$ & 90        & 16.122 & IV   \\
B14 & 1 & 0.602 & 0.2842 & 0.1874 & 0.2087 & $10^7$  & 1873.97 & 1.7783  & $1.778\times 10^{-4}$ & 100       & 16.349 & IV   \\
B15 & 1 & 0.716 & 0.3160 & 0.2189 & 0.2207 & $10^7$  & 2189.35 & 2.9907  & $2.991\times 10^{-4}$ & 200       & 18.469 & IV   \\
B16 & 1 & 0.901 & 0.3676 & 0.2679 & 0.2384 & $10^7$  & 2679.06 & 5.9460  & $5.946\times 10^{-4}$ & 500       & 22.416 & IV   \\
B17 & 1 & 1.071 & 0.3800 & 0.2796 & 0.2422 & $10^7$  & 2795.83 & 10.0000 & $1.000\times 10^{-3}$ & 1000      & 23.394 & IV   \\
B18 & 1 & 2.221 & 0.3834 & 0.2829 & 0.2430 & $10^7$  & 2828.79  & $\infty$& $\infty$             & $\infty$  & 23.829 & IV   \\
\hline
C1  & 1 & 0.107 & 0.0179 & 0.0076 & 0.0157 & $10^8$  & 240.30  & 0.0316  & $1.000\times 10^{-6}$ & 1         & 1.049  & I   \\
C2  & 1 & 0.127 & 0.0859 & 0.0138 & 0.0845 & $10^8$  & 436.94  & 0.0532  & $1.682\times 10^{-6}$ & 2         & 1.175  & II   \\
C3  & 1 & 0.152 & 0.1973 & 0.0320 & 0.1943 & $10^8$  & 1012.07 & 0.0894  & $2.828\times 10^{-6}$ & 4         & 2.005  & II   \\
C4  & 1 & 0.180 & 0.3398 & 0.0639 & 0.3334 & $10^8$  & 2019.92 & 0.1504  & $4.757\times 10^{-6}$ & 8         & 5.429  & III   \\
C5  & 1 & 0.191 & 0.3914 & 0.0715 & 0.3846 & $10^8$  & 2260.75 & 0.1778  & $5.623\times 10^{-6}$ & 10        & 7.009  & III   \\
C6  & 1 & 0.227 & 0.3113 & 0.0920 & 0.2970 & $10^8$  & 2909.05 & 0.2991  & $9.457\times 10^{-6}$ & 20        & 13.468 & III   \\
C7  & 1 & 0.251 & 0.2186 & 0.1013 & 0.1928 & $10^8$  & 3203.73 & 0.4054  & $1.282\times 10^{-5}$ & 30        & 16.929 & IV   \\
C8  & 1 & 0.269 & 0.1892 & 0.1158 & 0.1470 & $10^8$  & 3664.97 & 0.5030  & $1.591\times 10^{-5}$ & 40        & 20.934 & IV   \\
C9  & 1 & 0.285 & 0.1965 & 0.1198 & 0.1533 & $10^8$  & 3789.30 & 0.5946  & $1.880\times 10^{-5}$ & 50        & 22.953 & IV   \\
C10 & 1 & 0.298 & 0.2055 & 0.1256 & 0.1601 & $10^8$  & 3972.12 & 0.6817  & $2.156\times 10^{-5}$ & 60        & 25.011 & IV   \\
C11 & 1 & 0.310 & 0.2097 & 0.1286 & 0.1630 & $10^8$  & 4067.57 & 0.7653  & $2.420\times 10^{-5}$ & 70        & 26.264 & IV   \\
C12 & 1 & 0.320 & 0.2147 & 0.1331 & 0.1656 & $10^8$  & 4207.58 & 0.8459  & $2.675\times 10^{-5}$ & 80        & 27.665 & IV   \\
C13 & 1 & 0.330 & 0.2181 & 0.1368 & 0.1669 & $10^8$  & 4326.24 & 0.9240  & $2.922\times 10^{-5}$ & 90        & 28.536 & IV   \\
C14 & 1 & 0.339 & 0.2254 & 0.1421 & 0.1719 & $10^8$  & 4493.06 & 1.0000  & $3.162\times 10^{-5}$ & 100       & 29.816 & IV   \\
C15 & 1 & 0.403 & 0.2433 & 0.1588 & 0.1805 & $10^8$  & 5022.01 & 1.6818  & $5.318\times 10^{-5}$ & 200       & 33.136 & IV   \\
C16 & 1 & 0.506 & 0.2713 & 0.1886 & 0.1892 & $10^8$  & 5964.05 & 3.3437  & $1.057\times 10^{-4}$ & 500       & 36.253 & IV   \\
C17 & 1 & 0.602 & 0.3045 & 0.2209 & 0.1996 & $10^8$  & 6984.41 & 5.6234  & $1.778\times 10^{-4}$ & 1000      & 41.104 & IV   \\
C18 & 1 & 2.221 & 0.3168 & 0.2332 & 0.2018 & $10^8$  & 7372.93 & $\infty$& $\infty$              & $\infty$  & 43.733 & IV   \\
\enddata
\tablecomments{$\Gamma$ is the lateral-to-height aspect ratio; $\ell_{c}$ is the critical wavelength for the onset of convection; $u''$ is the rms velocity; $u_{z}''$ is the rms vertical velocity; $u_{h}''$ is the rms horizontal velocity; $Ra$ is the Rayleigh number; $Re_{z}$ is the vertical Reynolds number; $Ro$ is the convective Rossby number; $E$ is the Ekman number; $\widetilde{Ra}$ is the modified Rayleigh number; $Nu$ is the Nusselt number; and the cases are classified into four regimes according to flow patterns. Temporally and spatially (the whole box) averaged values are reported.}
\end{deluxetable*}

\section{Results}
\subsection{Flow pattern}
\begin{figure}[!htbp]
\plotone{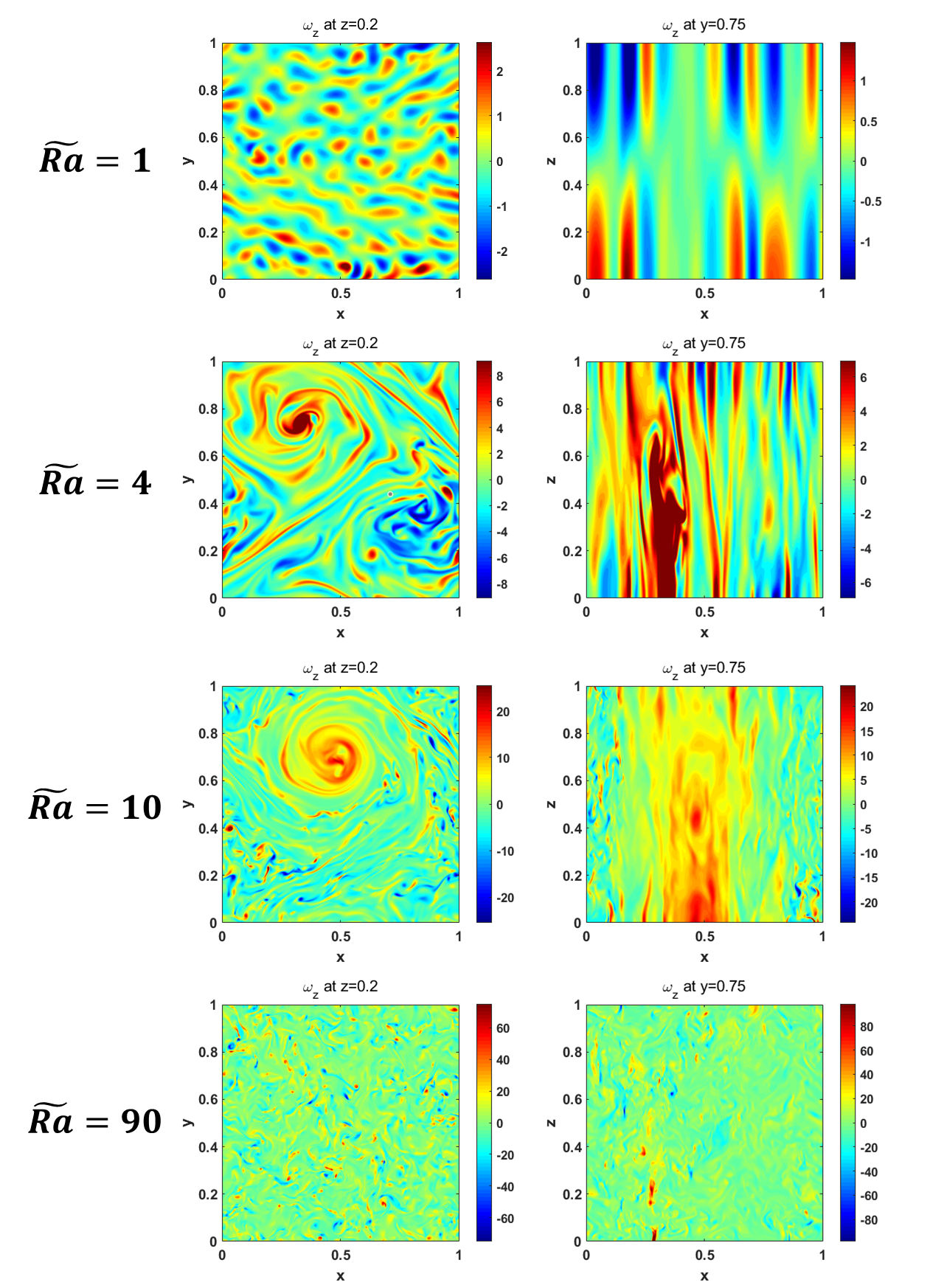}
\caption{The first column shows the horizontal cross-sections (at z=0.2) of the axial vorticity; the second row shows the vertical cross-sections (at y=0.75) of the axial vorticity. Red (blue) color denotes positive (negative) value. The Rayleigh number is $Ra=10^{8}$. The modified Rayleigh numbers are $\widetilde{Ra}=1,4,10,90$ from the top to bottom, respectively.}
\label{fig:fig1}
\end{figure}

Fig.~\ref{fig:fig1} shows the flow structures for the simulation cases with $Ra=10^8$. The left column displays the horizontal cuts of the axial vorticity at the plane $z=0.25$. The right column displays the vertical cuts of the axial vorticity at the plane $y=0.75$. Simulation cases with $\widetilde{Ra}=1,4,10$ and $90$ (cases C1, C3, C5, and C13) are arranged from the top to bottom. Obviously the flow structures are quite different for these cases. For the case $\widetilde{Ra}=1$, the flow is near the onset of oscillatory convection. The structure shows up as small vortices of alternately positive and negative axial vorticities. The structure of axial vorticity tends to be antisymmetric about the midplane. The vortices pattern is similar to those observed in simulations and experiments of \citet{chong2020vortices}, but the motions are different. In their simulations and experiments at $Pr=4.38$, they observed that small vortices moved as Brownian motions. However, in our simulations small vortices tend to move as following travelling waves (see animation for Fig.~\ref{fig:regimeI}). The reason for the difference is probably that we use a small Prandtl number, such that oscillatory wave can be easily excited in our cases. For the case $\widetilde{Ra}=4$, coexisted large-scale cyclone and anticyclone extend throughout the whole domain. Intense shear flow is driven in the interacting region of these LSVs. Time evolution of flow structure shows that these LSVs are formed through clustering and merging processes of small vortices (see animation for Fig.~\ref{fig:regimeII}). For the case $\widetilde{Ra}=10$, only a large-scale cyclone appears, accompanying with small vortices advected by a large-scale anticyclonic circulation (see animation for Fig.~\ref{fig:regimeIII}). For the case of $\widetilde{Ra}=90$, the condensation of convective flow disappears, and the flow structure tends to be three-dimensional turbulent (see animation for Fig.~\ref{fig:regimeIV}).

\begin{figure}[!htbp]
\plotone{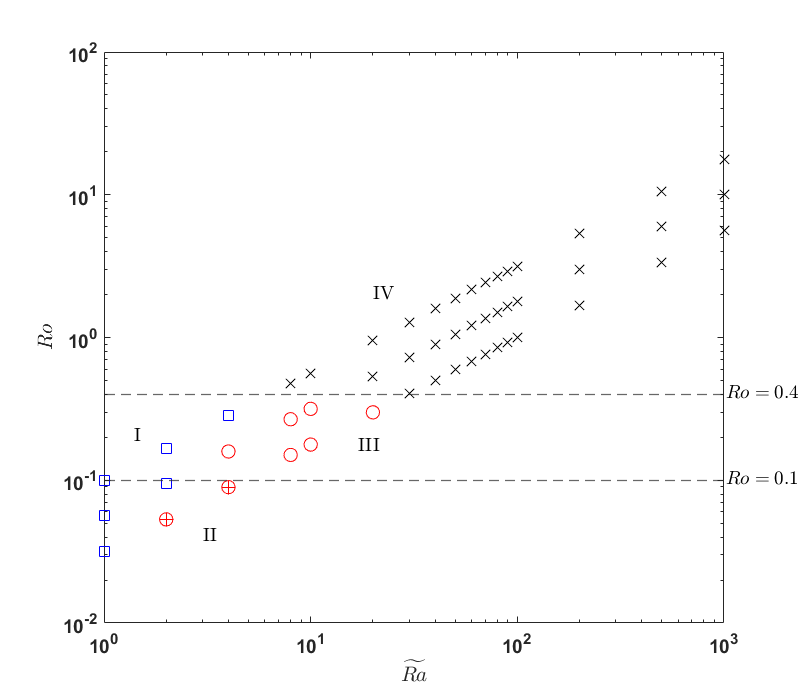}
\caption{Summary of simulation cases of rotating convection performed in this paper on the $\widetilde{Ra}-Ro$ plane. From the upper left to the bottom right are the three groups of simulations with different Rayleigh number $Ra=10^6$, $10^7$, and $10^8$. The symbols represent different convective behavioral regimes. Four different regimes are identified here as the square symbol for multiple small vortices, the circle plus symbol for coexisted large-scale cyclone and anticyclone, the circle symbol for large-scale cyclone, and the cross symbol for turbulence.}\label{fig:fig2}
\end{figure}

Fig.~\ref{fig:fig2} summarizes the simulation cases on the $\widetilde{Ra}-Ro$ plane. In this figure, three groups of simulation cases with different Rayleigh numbers $Ra=10^6$, $10^7$, and $10^8$ are marked as in lines with slopes of $3/4$ ($Ra$ increases from the upper left to the bottom right).  Four regimes are identified with different symbols on the figure: the square symbol for multiple small vortices (Regime I); the circle plus symbol for coexisted large-scale cyclone and anticyclone (Regime II); the circle symbol for large-scale cyclone (Regime III); and the cross symbol for turbulence (Regime IV). The appearance of large-scale vortices highly depends on the system-scale Rossby number $Ro$ and vertical Reynolds number $Re_{z}$. Here $Re_{z}$ is defined as $Re_{z}=u''_{z}Re$, where the double prime denotes root-mean-square (rms) average on the whole domain. The empirical result shows that two conditions must be satisfied for the emergence of large-scale vortices. First, the system-scale Rossby number must be smaller than a value of order unity. Our simulations suggest $Ro \lesssim 0.4$. Second, the vertical Reynolds number $Re_{z}$ should be larger than a value of about 400, so that turbulent convection can be developed in the vertical direction. Coriolis force plays an important role in the formation of large-scale vortices. When $Ro$ is large, Coriolis effect is unimportant and the flow is more likely to be turbulent if the Reynolds number is large enough. Large-scale cyclone appears when $Ro$ is small enough. The line $Ro=0.4$ in Fig.~\ref{fig:fig2} clearly separates Regime IV from Regime III. When $Ro$ further decreases, large-scale anticyclone emerges and coexists with large-scale cyclone. In our calculation, the transition from Regime III to Regime II occurs at around $Ro=0.1$. Apart from Regime II and Regime III, another regime with multiple small vortices (Regime I) exists when Rossby number is small. In Regime I, the Rayleigh number is just above the supercritical value. $Re_{z}$ is smaller than 400 in this regime, which means the flow is more likely to be laminar than turbulent along the vertical direction. The conditions on the appearance of LSVs were also discussed in \citet{favier2014inverse} and \citet{guervilly2014large}, but only for large-scale cyclones in their cases. They have identified that large-scale cyclones appear when Rossby number is smaller than a critical value and Reynolds number is greater than a value of about 100. Our simulation contributes to the literature of RBBC by showing that a regime of coexisted large-scale cyclones and anticyclones may appear when the $Ro$ further decreases. \citet{kapyla2011starspots} and \citet{chan2013numerical} also investigated the condition for the emergence of LSVs in their simulations on compressible convection. Apart from Regime II and III, they have reported another regime where anticylone dominates. However, this regime has not been identified in our current numerical result, probably because of the lack of compressible effect in the Boussinesq flow.

\subsection{Energy transfer}
To investigate how energy is distributed, we first calculate the power spectral density of kinetic energy at different wavenumbers. Since our simulation cases are aperiodic in the vertical direction, we only compute the two-dimensional kinetic energy spectrum $P_{2}(k)$ on the horizontal space \citep{chan1996turbulent,cai2018numerical}. For a Boussinesq flow, the kinetic energy density $P_{2}(k)$ at a specific layer can be evaluated as \citep{cai2020penetrative}
\begin{eqnarray}
\int P_{2}(k)dk=\sum_{m}\sum_{n} a_{m,n}^2 (|\Psi_{m,n}|^2+|D_{z}\Phi_{m,n}|^2+a_{m,n}^2|\Phi_{m,n}|^2)~,
\end{eqnarray}
where $k=[(m^2+n^2)^{1/2}]$ is the horizontal wavenumber (the brackets mean the number is round off to an integer), $a_{m,n}=(2\pi/L) (m^2+n^2)^{1/2}$, $L=1$ is the lateral size of the box, and $m,n$ are the spectral numbers in the $x$ and $y$ directions, respectively.

\begin{figure}[!htbp]
\plotone{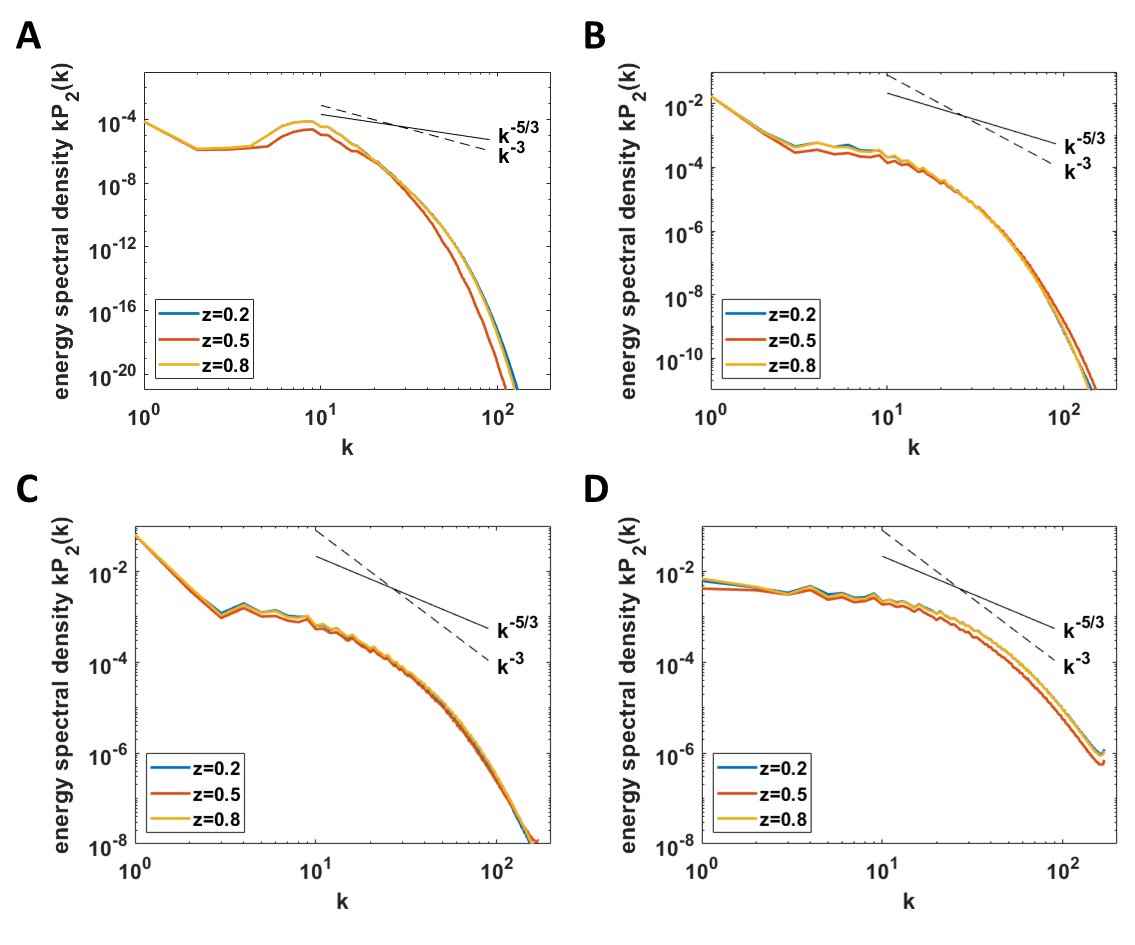}
\caption{(A-D)Compensated power spectral density of kinetic energy $kP_{2}(k)$ versus horizontal wavenumber $k$ for cases C1 ($\widetilde{Ra}=1$), C3 ($\widetilde{Ra}=4$), C5 ($\widetilde{Ra}=10$), and C13 ($\widetilde{Ra}=90$), respectively. Power spectral densities at different layers $z=0.1,0.5,0.9$ are shown with blue, red, and yellow colors, respectively. Scalings of $k^{-5/3}$ (Kolmogrov inertial power law) and $k^{-3}$ (scaling of 2D enstrophy cascade) are shown with solid and dashed lines for references.\label{fig:fig3}}
\end{figure}

Fig.~\ref{fig:fig3} shows the compensated power spectral density $kP_{2}(k)$ as a function of $k$ for cases C1 ($\widetilde{Ra}=1$), C3 ($\widetilde{Ra}=4$), C5 ($\widetilde{Ra}=10$), and C13 ($\widetilde{Ra}=90$), respectively. First, we see that the power spectral densities do not vary significantly at different heights, despite the fact that a faster decay rate is observed for small scale motions at the midplane in rapidly rotating cases. Second, we notice that the energy spectral densities scale approximately as $k^{-3}$ within the wavenumber range $1\leq k \leq 3$ for cases C3 and C5, which is consistent with the scaling in a large-scale condensation.


In order to study the effects of rotation on the energy transfer of turbulent convection, we decompose the fluid motion into depth-averaged barotropic (2D) and depth-dependent baroclinic (3D) components \citep{julien2012statistical,favier2014inverse}. For example, the 2D barotropic and 3D baroclinic components of velocity are defined as
\begin{eqnarray}
\langle \boldsymbol{u} \rangle(t,x,y)=\int_{0}^{1}\boldsymbol{u}(t,x,y,z)dz~,\\
\boldsymbol{u}'(t,x,y,z)=\boldsymbol{u}(t,x,y,z)-\langle \boldsymbol{u} \rangle(t,x,y)~,
\end{eqnarray}
respectively. Also, we use the symbol overbar (e.g. $\overline{\langle u_{z} \rangle}$) to represent the corresponding temporal averages.

\begin{figure}[!htbp]
\plotone{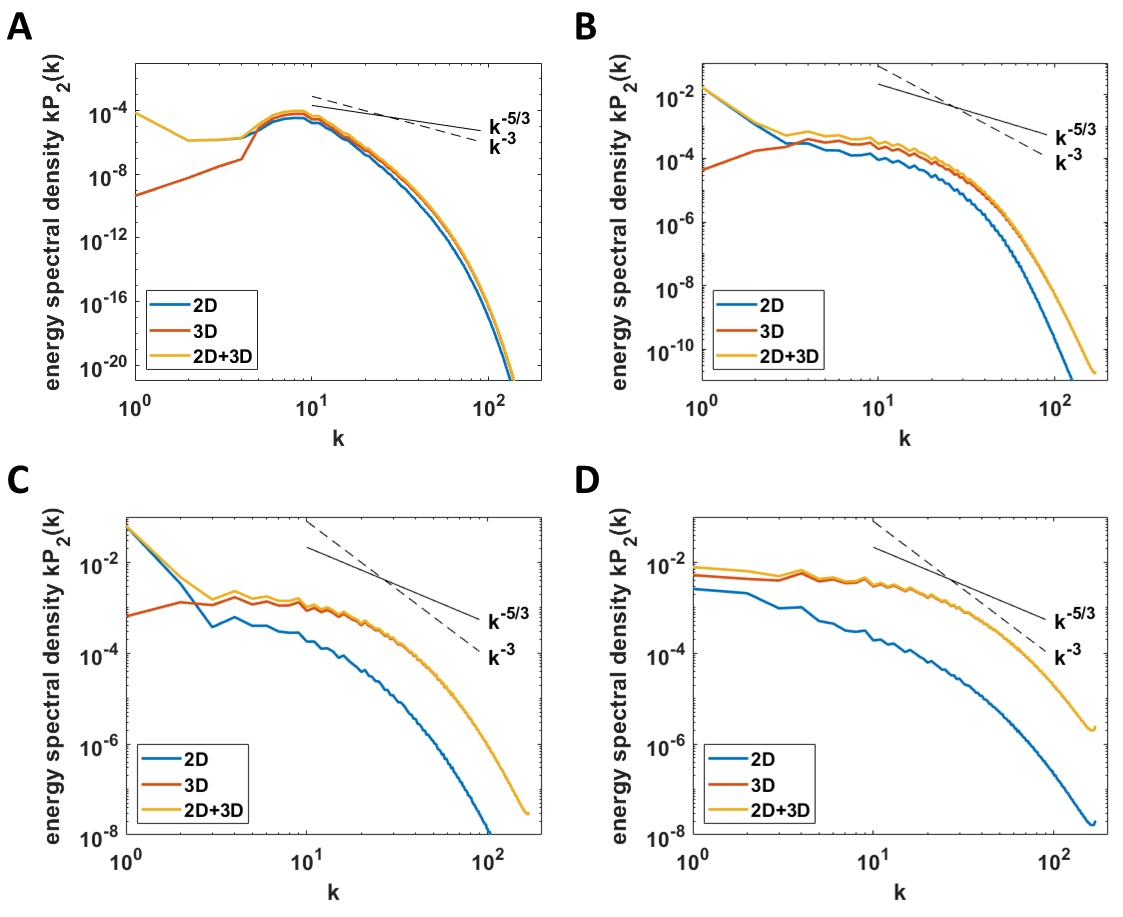}
\caption{(A-D)Compensated averaged power spectral density of kinetic energy $kP_{2}(k)$ versus horizontal wavenumber $k$ for cases C1 ($\widetilde{Ra}=1$), C3 ($\widetilde{Ra}=4$), C5 ($\widetilde{Ra}=10$), and C13 ($\widetilde{Ra}=90$), respectively. Power spectral densities of 2D, 3D components and their summation are shown with blue, red, and yellow colors, respectively. Scalings of $k^{-5/3}$ (Kolmogrov inertial power law) and $k^{-3}$ (scaling of 2D enstrophy cascade) are shown with solid and dashed lines for references.\label{fig:fig4}}
\end{figure}

After the decomposition, we can calculate the power spectral densities of the kinetic energies of 2D barotropic and 3D baroclinic components. Fig.~\ref{fig:fig4} presents the vertically averaged power spectral density of $kP_{2}(k)$ as functions of $k$. The contributions from 2D and 3D components and their summation are shown with blue, brown, and yellow lines, respectively. The Kolmogrov $-5/3$ scaling law in three-dimensional turbulence and the enstrophy cascade $-3$ scaling law in two-dimensional turbulence are also shown for references. For the case $\widetilde{Ra}=90$ shown in Fig.~\ref{fig:fig4}D, the energy contained in the 2D component is small compared to that in 3D component. It is reasonable since in this regime the rotational effect is small and the flow is more likely to be three-dimensional turbulent. The 2D effect starts to play a role, when rotation rate increases and LSVs appear (cases $\widetilde{Ra}=4$ and $\widetilde{Ra}=10$). Fig.~\ref{fig:fig4}B and \ref{fig:fig4}C clearly shows that more energy is contained in 2D component for large-scale motions (small $k$).
For small-scale motions, more energy is still contained in 3D component. When $\widetilde{Ra}$ further decreases to 1 (Fig.~\ref{fig:fig4}A), the energy contained in the 2D component is comparable to that in the 3D component for small-scale motions, and much larger than that in the 3D component for large-scale motions. From the discussion, we see that the 2D effect are more and more important when $Ro$ decreases.

\begin{figure}[!htbp]
\plotone{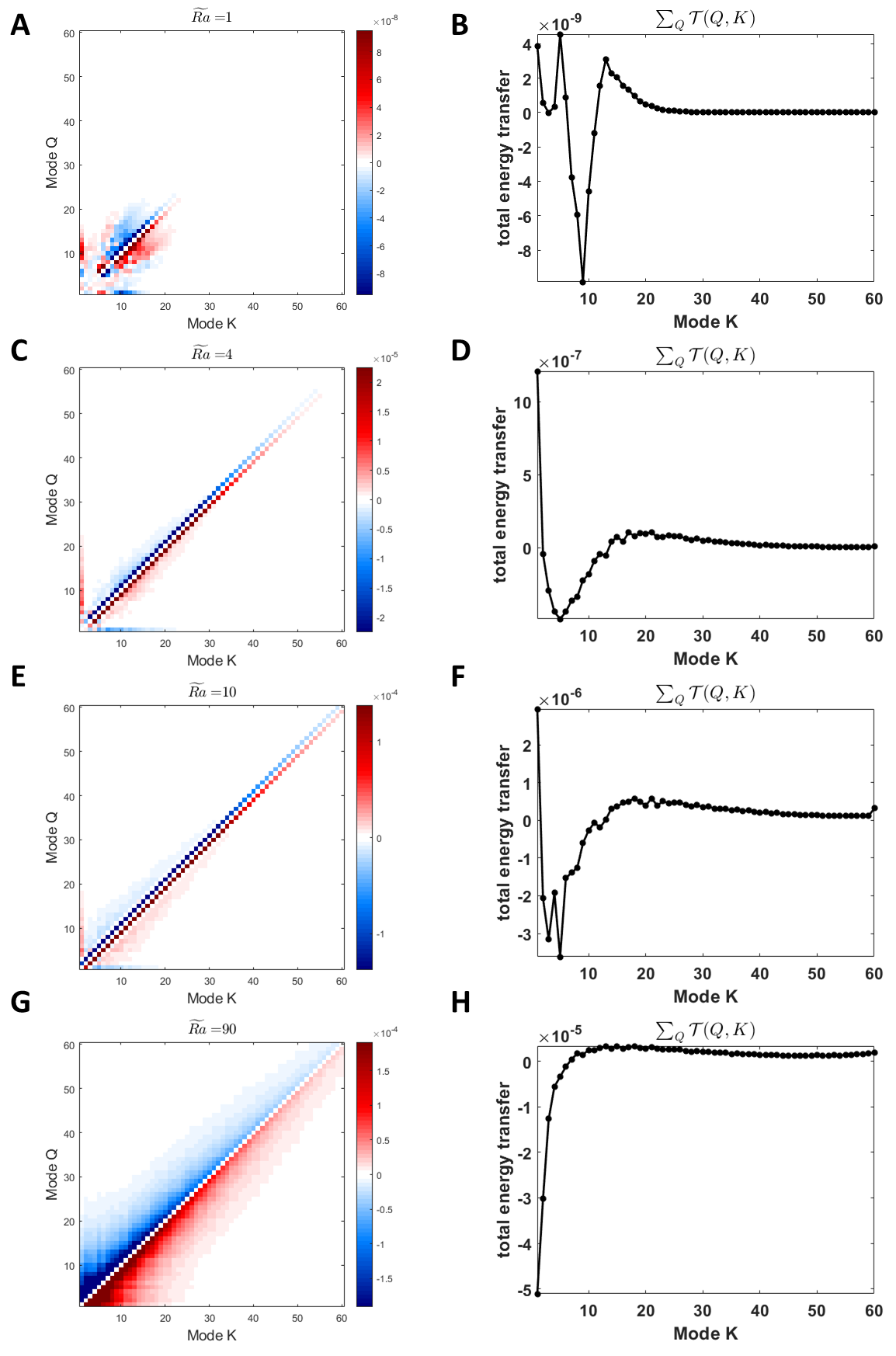}
\caption{The left column shows the heatmaps of the kinetic energy transfer coefficients $\mathcal{T}(Q,K)$ for shells $1\leq Q,K\leq 60$. Positive (negative) value means that energy is taken from (injected to) mode Q. The Rayleigh number is $Ra=10^{8}$. Note that values close to zero have been removed from the heatmap to signify the major energy transfer processes. The averaged period is about 10 units of time. The right column shows the corresponding values of $\sum_{Q}\mathcal{T}(Q,K)$ for the left column.\label{fig:fig5}}
\end{figure}

To further illustrate the transfer of kinetic energy among different scales, we define the kinetic energy transfer coefficients from shell Q to shell K \citep{alexakis2005shell,mininni2005shell,favier2014inverse} as
\begin{equation}
\mathcal{T}(Q,K)=-\int_{V}\overline{\boldsymbol{u}_{K}\cdot (\boldsymbol{u}\cdot \nabla \boldsymbol{u}_{Q})}dV~,
\end{equation}
where the shells $Q$ and $K$ are defined on the spectral space with horizontal wave numbers in the range of $k\in(Q-1,Q]$ and $k\in(K-1,K]$, respectively; $u_{Q}$ and $u_{K}$ are the corresponding shell filtered velocities; and $V$ is the volume of computational domain. Positive (negative) $\mathcal{T}(Q,K)$ means that the kinetic energy is taken from (given to) shell $Q$ and given to (taken from) shell $K$. The left column of Fig.~\ref{fig:fig5} shows $\mathcal{T}(Q,K)$ four selected cases in the four regimes. For the case $\widetilde{Ra}=90$ in the regime IV, the flow is not condensed and the kinetic energy is taken from larger scales to smaller scales. However, for the case $\widetilde{Ra}=10$ in the regime III, we see from Fig.~\ref{fig:fig5} that energy can be directly transferred from small scales ($K\leq 18$) to the largest scale. This nonlocal inverse energy cascade provides energy to sustain the formation of the large-scale cyclone. It is in agreement with the results of the previous numerical simulations on RRBC \citep{favier2014inverse,guervilly2014large}. For the case $\widetilde{Ra}=4$ in the regime II, despite that energy is taken from small scales ($5\leq K\leq 22$), we also observe a direct energy transfer from the largest scale to moderate scales ($K=4$). The reason might be associated with the appearance of anticyclone. Compared with cyclone, the convection is more turbulent within the anticyclone because the effective rotating speed is smaller \citep{chan2013numerical,kapyla2011starspots}. With more vigorous turbulent motions, the energy is more likely transferred from large scales to small scales. For the case $\widetilde{Ra}=1$ in the regime I, the vertical Reynolds number is small and the Ekman number is high, and the rotation effect dominates the buoyant effect at small scales. Thus we observe that the kinetic energy transfer is cut off at small scales ($K\leq 17$). The right column of Fig.~\ref{fig:fig5} shows $\sum_{Q}\mathcal{T}(Q,K)$, which is the total energy transferred from all $Q$ to a single shell $K$. A negative value of $\sum_{Q}\mathcal{T}(Q,K)$ means that the wavenumber $K$ supplies energy, thus this wavenumber can be thought as a forcing wavenumber. From the right column of Fig.~\ref{fig:fig5}, we see that the forcing wavenumbers for cases C1, C3, C5, and C13, are $7\leq K \leq 11$, $2\leq K \leq 13$, $2\leq K \leq 12$, and $1\leq K \leq 6$, respectively. When rotational effect increases, the forcing wavenumbers tend to shift from large-scale wavenumbers to small-scale wavenumbers. In the turbulence regime (regime IV), the mode $K=1$ is a forcing wavenumber. However, it is no longer a forcing wavenumber in other regimes where the rotational effects are dominant.

\begin{figure}[!htbp]
\plotone{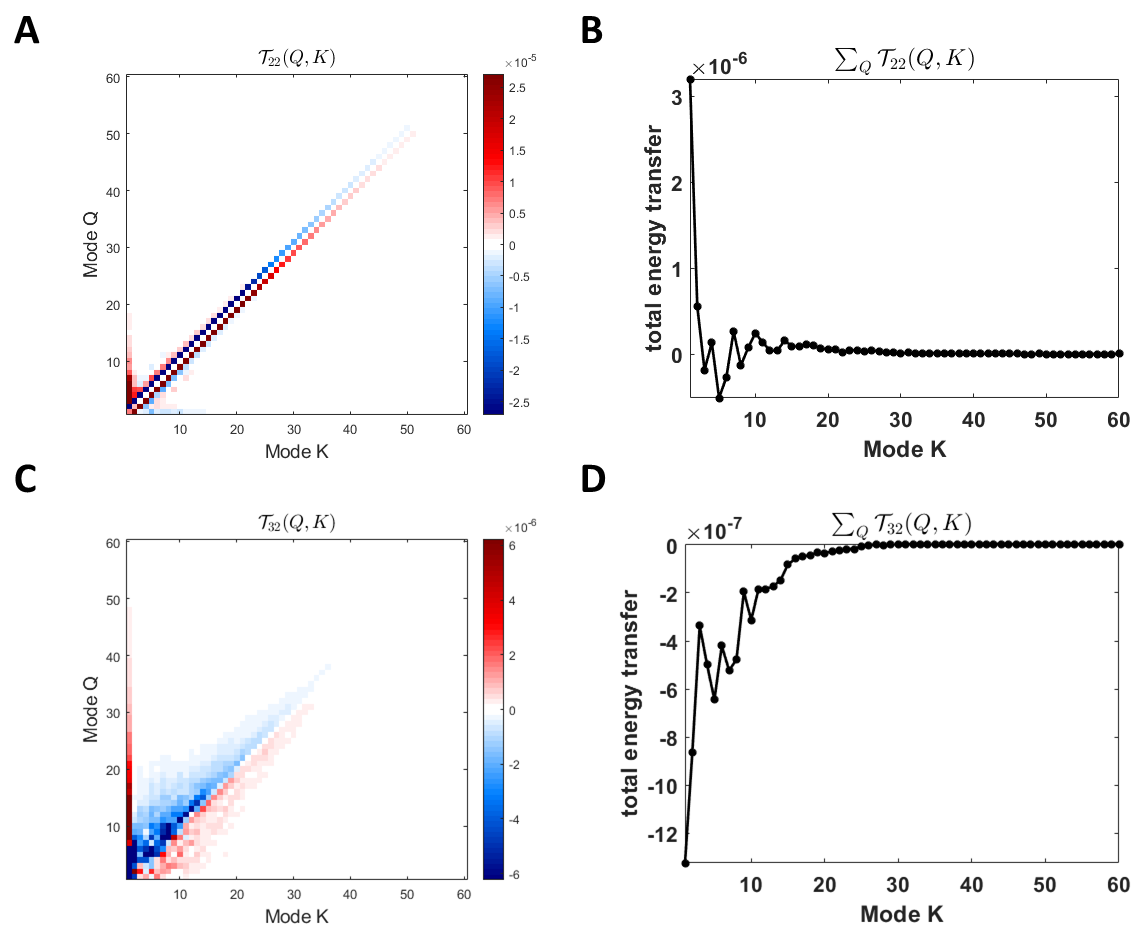}
\caption{The left column shows the heatmaps of $\mathcal{T}_{22}(Q,K)$ and $\mathcal{T}_{32}(Q,K)$ for shells $1\leq Q,K\leq 60$ of case C5. Positive (negative) value means that energy is taken from (injected to) mode Q. Note that values close to zero have been removed from the heatmap to signify the major energy transfer processes. The averaged period is about 10 units of time. The right column shows the total energy transferred from the 2D and 3D components of shell $Q$ to the 2D component of shell $K$, defined as $\sum_{Q}\mathcal{T}_{22}(Q,K)$ and $\sum_{Q}\mathcal{T}_{32}(Q,K)$, respectively.\label{fig:fig6}}
\end{figure}

To consider the energy transfer between the 2D and 3D components, we further define the the self- and cross-transfer coefficients as
\begin{eqnarray}
\mathcal{T}_{22}(Q,K)&=&-\int_{V}\overline{\langle \boldsymbol{u}_{K} \rangle\cdot(\boldsymbol{u} \nabla\cdot \langle \boldsymbol{u}_{Q} \rangle)}dV~,\\
\mathcal{T}_{32}(Q,K)&=&-\int_{V}\overline{\langle \boldsymbol{u}_{K} \rangle\cdot(\boldsymbol{u} \nabla\cdot \boldsymbol{u}_{Q}' )}dV~,
\end{eqnarray}
where $\mathcal{T}_{22}(Q,K)$ and $\mathcal{T}_{32}(Q,K)$ measure the interactions of the 2D and 3D components of shell $Q$ with the 2D component of shell $K$, respectively. Figs.~\ref{fig:fig6}A and \ref{fig:fig6}C show the heatmaps of $\mathcal{T}_{22}(Q,K)$ and $\mathcal{T}_{32}(Q,K)$ for case C5. From Fig.~\ref{fig:fig6}A, we note that the local energy transfer from large to small scales is an important process in self-interaction of 2D components. Apart from the local energy transfer, we also observe a significant direct energy transfer process from moderate scales $3 \leq Q \leq 14$ to the largest scale $K=1$. The cross interaction of 3D and 2D components (Fig.~\ref{fig:fig6}C), however, shows that the moderate scales of 3D component ($2\leq Q \leq 6$) take energy from the largest scale of 2D component ($K=1$). The small scales of 3D component ($8\leq Q \leq 48$), on the other hand, put energy into the largest scale of 2D component ($K=1$). To investigate the net effect of self- and cross-transfer of kinetic energy, we have computed the total energy transfer from all shell $Q$ to a single shell $K$. Figs.~\ref{fig:fig6}B  and \ref{fig:fig6}D show $\sum_{Q}\mathcal{T}_{22}(Q,K)$ and $\sum_{Q}\mathcal{T}_{32}(Q,K)$ as a function of $K$, respectively. Apparently, the 3D component has a net effect of taking energy from shell $K=1$ (Fig.~\ref{fig:fig6}D). However, the 2D component shows a net effect of putting energy into the shell $K=1$ (Fig.~\ref{fig:fig6}B), indicating that the large-scale vortices are probably maintained by a 2D self-transfer process in this case.

\subsection{Heat transfer}
In this subsection, we discuss the heat transfer in RBBC. To study the 2D and 3D effects on heat transfer, we separate the heat flux into 2D and 3D components. To achieve this, we first split $u_{z}$ and $\Theta$ into two parts, by letting $u_{z}=\langle u_{z}\rangle+ u_{z}'$ and $\Theta=\langle\Theta\rangle + \Theta'$. Here $u_{z}'$ and $\Theta$ are three-dimensional perturbations from the two-dimensional integrated mean values. Then we can split the convective flux into 2D and 3D components by
\begin{eqnarray}
F_{c}=Pe\overline{\langle u_{z}\Theta\rangle}=F_{c,2D}+F_{c,3D},
\end{eqnarray}
where
\begin{eqnarray}
F_{c,2D}(x,y)&=&Pe\overline{\langle u_{z}\rangle\langle\Theta\rangle}~,\\
F_{c,3D}(x,y)&=&Pe\overline{\langle u'_{z}\Theta'\rangle}~.
\end{eqnarray}
Here $F_{c,2D}$ measures the heat flux transported by 2D convection, and $F_{c,3D}$ measures the heat flux transported by 3D convection. Both $F_{c,2D}$ and $F_{c,3D}$ take average on convective flux temporally and vertically, and thus they are functions of $x$ and $y$. As mentioned earlier, both cyclonic and anticyclonic regions appear in regimes II and III. The cyclonic and anticyclonic regions can be shown more clearly after taking temporal and vertical averages. As were shown in the animations for Figs.~\ref{fig:regimeII} and Fig.~\ref{fig:regimeIII}, the cyclonic and anticyclonic regions move around with time, therefore the temporal average cannot be taken for too long a period. For both cases, we take the average for a period of about 20 units of time. The nondimensional system rotation period is $4\pi Ro$, thus the averaged time covers about 18 system rotation periods for case C3 and 9 system rotation periods for case C5. Within this time period, the cyclones and anticyclones do not drift too much away.

\begin{figure}[!htbp]
\plotone{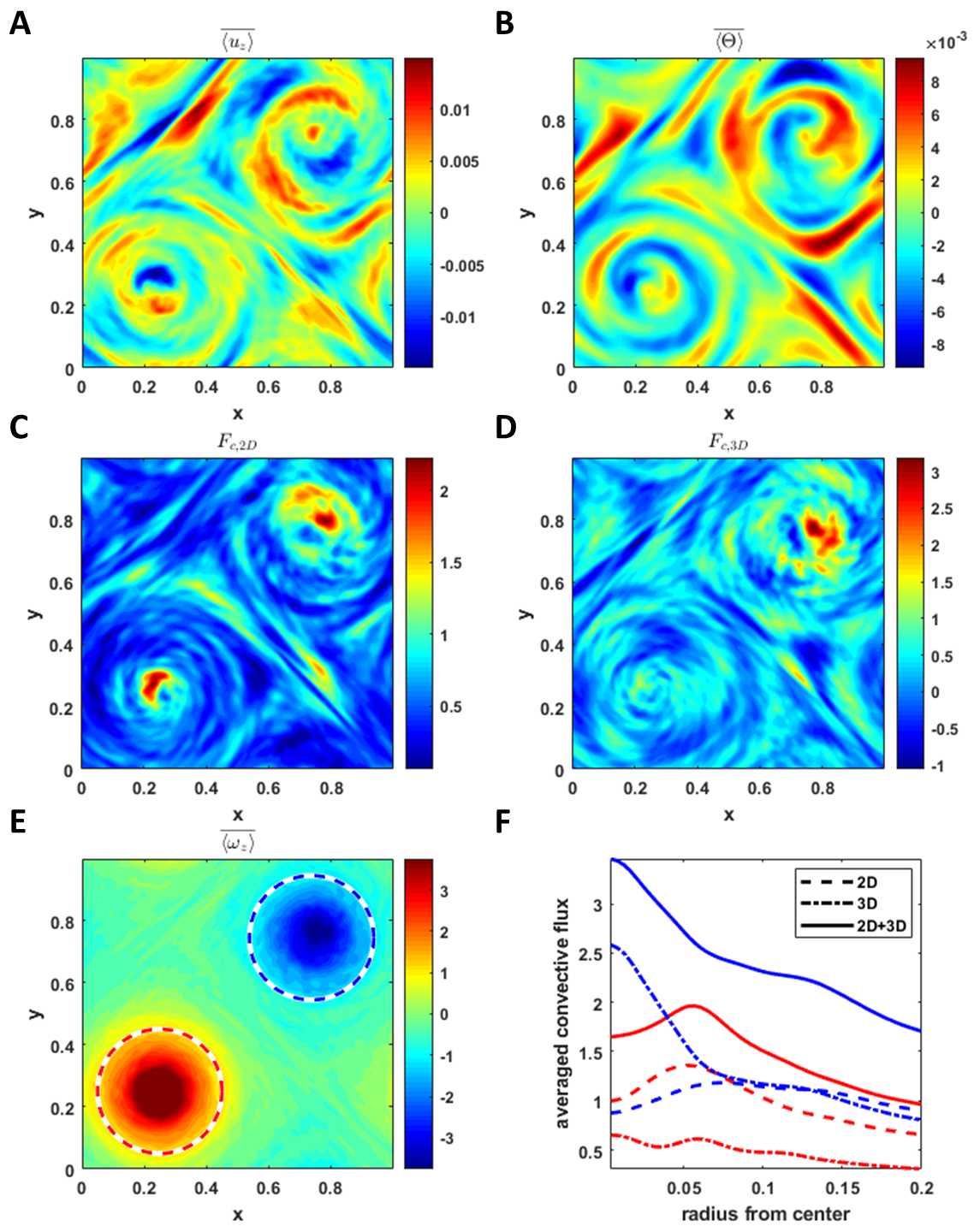}
\caption{The first five panels in sequence show vertically averaged vertical velocity, temperature perturbation, convective flux transported by 2D barotropic component, convective flux transported by 3D baroclinic component, and vertical component of vorticity. The cyclone and anticyclone (or anticyclonic region) are shown by red and blue circles in the fifth panel (bottom left), respectively. The last panel shows the averaged convective flux in cyclone (red curves) and anticyclone (or anticyclonic region, blue curves) as functions of radius from their respective center. The average is taken temporally, vertically, and horizontally within the circles plotted in the fifth panel. The case is C3 with $Ra=10^{8}$ and $\widetilde{Ra}=4$. The temporal average period is about 20 units of time.\label{fig:fig7}}.
\end{figure}

\begin{figure}[!htbp]
\plotone{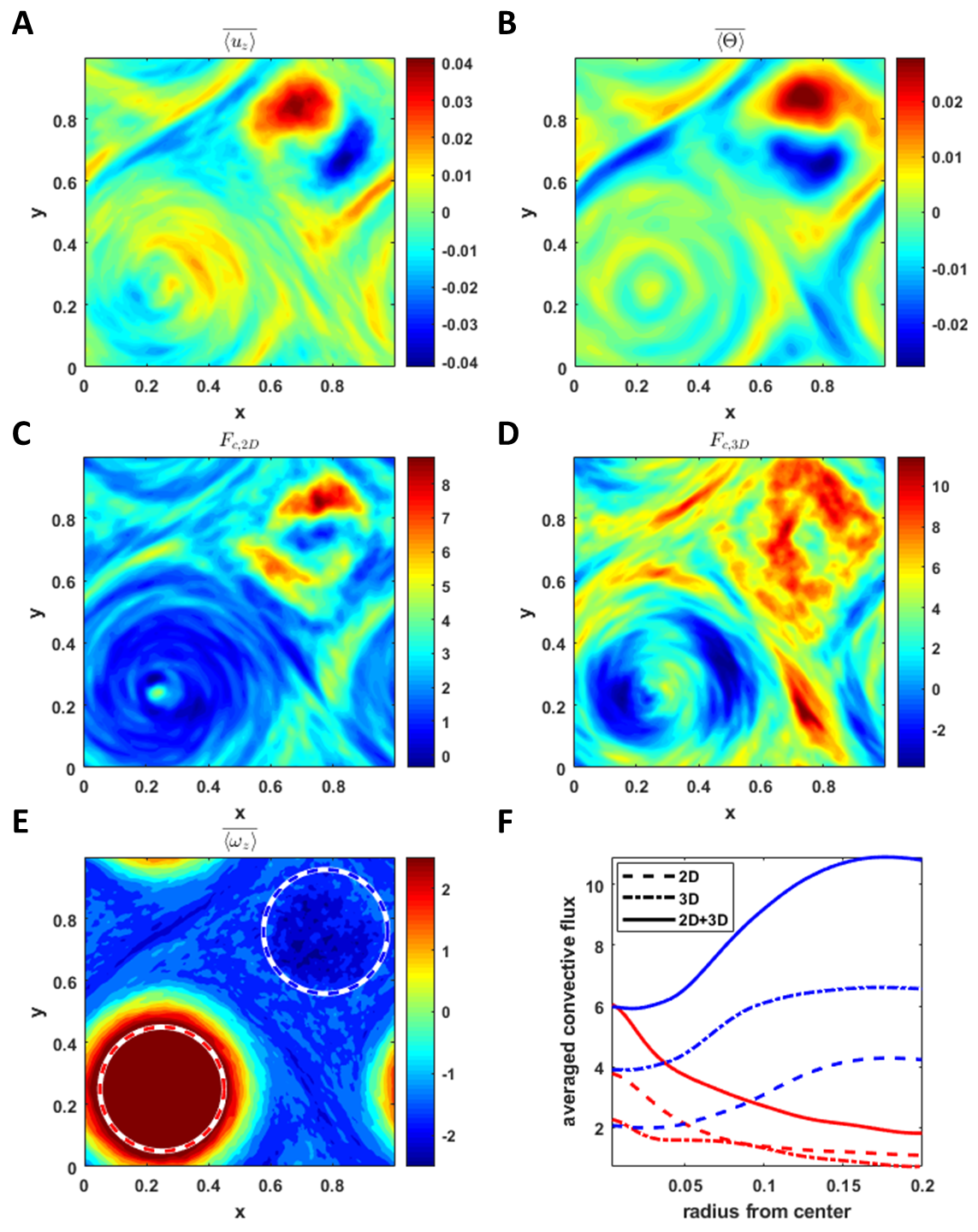}
\caption{Companion to Fig.\ref{fig:fig7}, but for the case C5 with $Ra=10^{8}$ and $\widetilde{Ra}=10$\label{fig:fig8}}.
\end{figure}

The results of case C3 (Regime II) and C5 (Regime III) are shown in Fig.~\ref{fig:fig7} and Fig.~\ref{fig:fig8}, respectively. The pattern at Fig.~\ref{fig:fig7}E shows clearly the cyclonic (the region marked by red circle) and anticyclonic (the region marked by blue circle) regions on the contour plot of the averaged vertical component of vortical structure $\overline{\langle\omega_{z}\rangle}$. Figs.~\ref{fig:fig7}A and \ref{fig:fig7}B show the averaged vertical velocity $\overline{\langle u_{z}\rangle}$ and temperature perturbation $\overline{\langle\Theta\rangle}$. Both $\overline{\langle u_{z}\rangle}$ and $\overline{\langle\Theta\rangle}$ show structures of twisted rolls in the cyclonic and anticyclonic regions. In simulations of rapidly rotating compressible flow, \citet{chan2013numerical} have found that a cyclone has lower temperature, while an anticyclone has higher temperature in the core region. Our result of RBBC shows different temperature structures in these LSVs. One possible reason may be that our RBBC simulations lack of compressible effect. In Boussinesq flow, the effect of density variation is only considered in the buoyancy term, while the variation on the horizonal direction is ignored. However, in the compressible flow, the horizonal variation of density has significant effect on the horizontal variation of temperature, which may help create a lower (higher) temperature core for cyclone (anticyclone). Figs .~\ref{fig:fig7}C and \ref{fig:fig7}D show the convective fluxes transported by 2D and 3D components. We note that the 2D component $F_{c,2D}$ almost transports positive convective flux. The 3D component $F_{c,3D}$ tends to transport positive convective fluxes in the anticyclonic region. However, in the cyclonic region, the convective flux transported by it can either be positive or negative. It indicates the 3D turbulent motions plays more important role in transporting convective flux in the anticyclonic region rather than the cyclonic regions. Similar behaviours are observed in Fig.~\ref{fig:fig8}.

\begin{figure}[!htbp]
\plotone{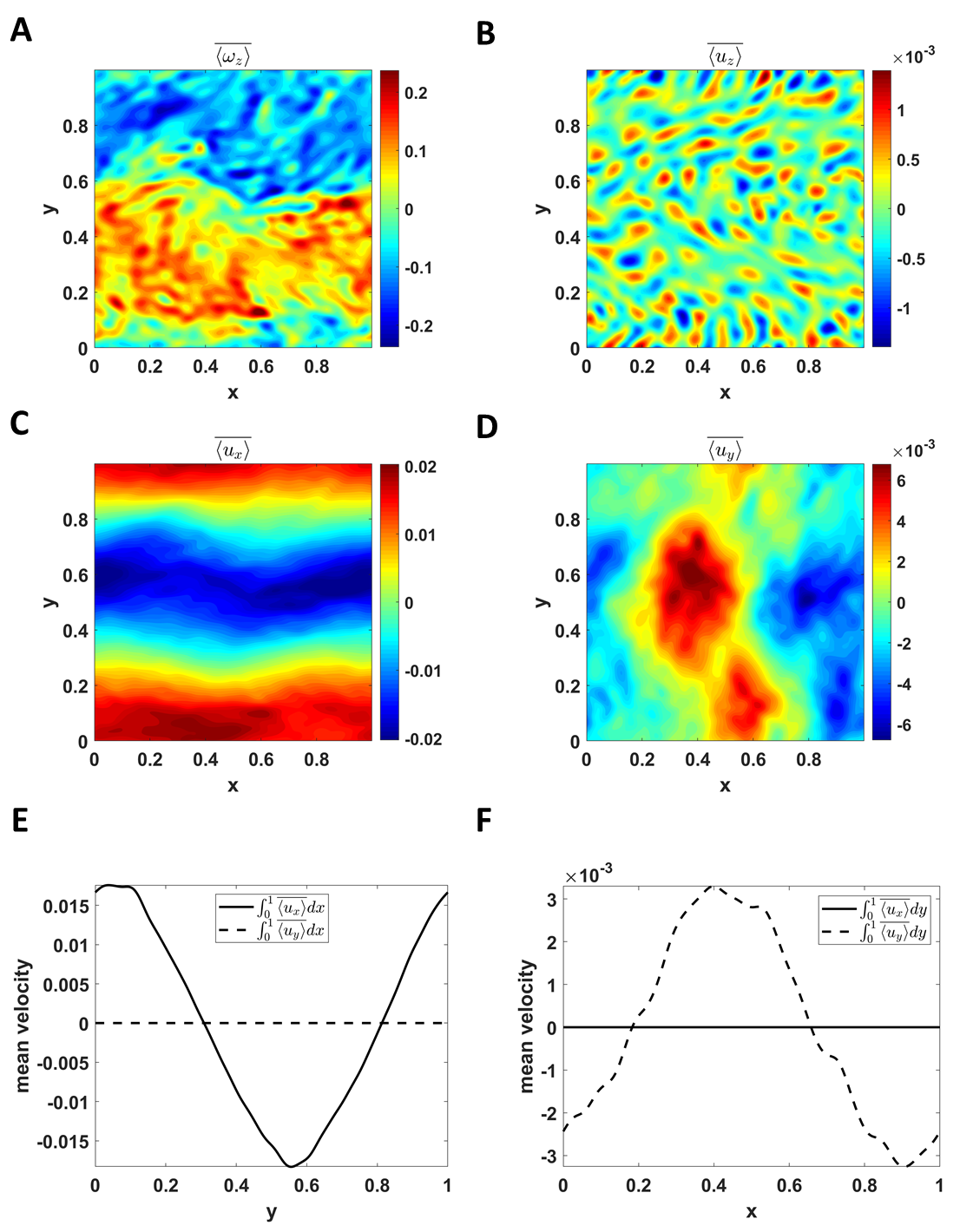}
\caption{Averaged values for case C1 with $Ra=10^{8}$ and $\widetilde{Ra}=1$. (A-D)Contour plots of the averaged vertical component of vorticity, vertical velocity, horizontal velocities along $x$ and $y$ directions, respectively. (E-F)Shear flow velocities as functions of $y$ and $x$, respectively.\label{fig:fig9}}.
\end{figure}

In order to quantitatively compare the efficiencies of heat transportation between the cyclonic and anticyclonic regions, we take averages on $F_{c,2D}$ and $F_{c,3D}$ within the disks around the cyclonic and anticyclonic spots (the centers of the red and blue circles in $\overline{\langle\omega_{z}\rangle}$), respectively. Fig.~\ref{fig:fig7}F shows the averaged convective fluxes in the cyclonic and anticyclonic regions for case C3 by red and blue curves, respectively. In the anticyclonic region, the contributions of heat transportation by 2D and 3D components are comparable. However, in the cyclonic region, the heat transportation by 2D component dominates that by 3D component. For case C5, the result is different. In the cyclonic region, the heat transportation by 2D and 3D components are comparable. On the other hand, in the anticyclonic region, the heat transportation by 3D component is much higher than that by 2D component. By comparing cases C3 and C5, we see that 2D component plays more and more important role in both cyclonic and anticyclonic regions when rotation rate increases. To verify this trend, we have also computed the averaged convective fluxes transported by 2D and 3D components in the case C1. It has been found that the convective flux transported by 2D component is about four times of that by 3D component. From the plot of $\overline{\langle\omega_{z}\rangle}$ in Fig.~\ref{fig:fig9}A, we see that case C1 also contains cyclonic and anticyclonic regions, but we have not found significant difference of convective fluxes transported between these two regions. Although the cyclonic and anticyclonic regions are well separated in an averaged sense (Fig.~\ref{fig:fig9}A), the averaged vertical velocity $\overline{\langle v_{z}\rangle}$ has not shown similar distribution (Fig.~\ref{fig:fig9}B). Apart from $\overline{\langle v_{z}\rangle}$, both the averaged horizontal velocities $\overline{\langle v_{x}\rangle}$ (Fig.~\ref{fig:fig9}C and \ref{fig:fig9}E) and $\overline{\langle v_{y}\rangle}$ (Fig.~\ref{fig:fig9}D and \ref{fig:fig9}F) have developed shear structures. The shear velocity in the $x$-direction is larger than that in the $y$-direction by an order of magnitude. As a result, the group motions of small vortices are more prominent along the $x$-direction (see animation for Fig.~\ref{fig:regimeI}). If the shear velocities in the $x$- and $y$-directions are comparable, then we would expect that large-scale vortices could be formed. To examine whether the shear flow is preferred in the $x$-direction somehow in the system or just by chance, we continue the simulation of case C1 by switching the $x$- and $y$-directions. From the animation for Fig.~\ref{fig:rot_regimeI}, we see that the shear flow is preferred in one direction by chance.

From the above discussion, we have the following conclusions. First, as rotation rate increases, 2D effect increases and 3D effect decreases in transporting convective flux. Second, in regimes II and III when LSVs appear, heat transfer by convection is more efficient in anticyclonic region than cyclonic region. It is consistent with the observation that convection in anticyclone is more turbulent because the effective rotation is smaller \citep{chan2013numerical}. In the simulations of Boussinesq flow, \citet{guervilly2014large} also observed a significant reduction of heat transfer inside the cyclone.

\subsection{Statistical results}
\begin{figure}[!htbp]
\plotone{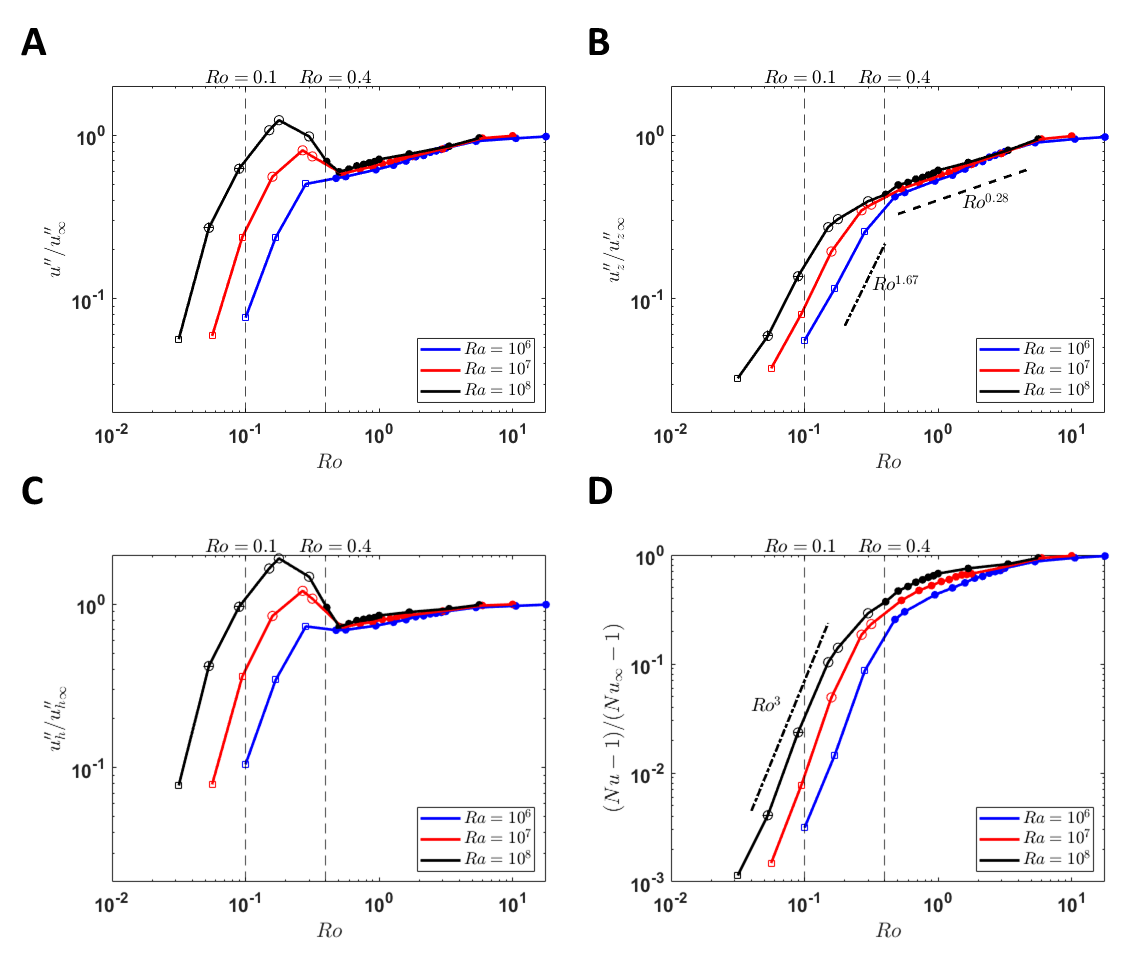}
\caption{Statistical results of normalized velocities and modified Nusselt number as functions of the convective Rossby number. (A) The total velocity $\langle u''\rangle$; (B) The vertical velocity $\langle u''_{z}\rangle$; (C) The horizontal velocity $\langle u''_{h}\rangle$; (D) The modified Nusselt number $(Nu-1)$. The results are normalized by the reference values of corresponding reference cases. The symbols dot, circle, circle plus, and square represent regimes I-IV, respectively. \label{fig:fig10}}.
\end{figure}

In this subsection, we investigate the effects of rotation on statistical results of velocities and Nusselt number. To compare the results among different groups, we use the non-rotating case within each group as a reference case and normalize all the statistical values by the corresponding reference values. Figs.~\ref{fig:fig10}A-\ref{fig:fig10}C show the normalized rms velocities $u''/u''_{\infty}$, $u''_{z}/{u''_{z}}_{\infty}$, and $u''_{h}/{u''_{h}}_{\infty}$ as functions of $Ro$, respectively. The subscript $\infty$ denotes the value of the reference non-rotating case. First, we note that the variation of $u''/u''_{\infty}$ on $Ro$ is not monotonic. It shows an increasing trend with increasing $Ro$ in the regime I, II, and IV, while a decreasing trend with increasing $Ro$ in the regime III. The trend reversal in the regime III is due to the appearance of large-scale cyclones, in which the flow is dominant by the horizontal large-scale motions. This can be further verified by looking at Figs.~\ref{fig:fig10}B and \ref{fig:fig10}C, where the reversal trend is only observed in the horizontal but not in the vertical velocity profiles. Second, we note that the normalized velocities in different groups almost collapse into a single curve in the regime IV. In this regime, the normalized velocities are mainly affected by the convective Rossby number. In the vicinity of $Ro\sim 1$ of regime IV, the normalized vertical velocity is estimated to approximately obey a scaling of $u''_{z}/{u''_{z}}_{\infty} \propto Ro^{2/7}$. In other fast rotating regimes, apart from the convective Rossby number, the normalized velocities also depend on Rayleigh numbers. Cases with higher Rayleigh number tend to have higher normalized vertical velocities. Although the curves of $u''_{z}/{u''_{z}}_{\infty}$ for different groups are separated, we still find that $u''_{z}/{u''_{z}}_{\infty}$ approximately obeys a scaling of $u''_{z}/{u''_{z}}_{\infty} \propto Ro^{5/3}$ in these regimes. Fig.~\ref{fig:fig10}D shows the normalized modified Nusselt number $(Nu-1)/(Nu_{\infty}-1)$, where the Nusselt number is defined as $Nu=\int \left[ Pe \langle \overline{u_{z}\Theta}\rangle - \langle \partial \overline{\Theta}/{\partial z} \rangle \right] dz +1$. As seen from the figure, $(Nu-1)/(Nu_{\infty}-1)$ always increases with increasing $Ro$, which indicates that rotation has an negative effect on heat transfer. The slope of $(Nu-1)/(Nu_{\infty}-1)$ has a decreasing trend with increasing $Ro$. In the rapidly rotating regimes ($Ro<0.4$), the curves of $(Nu-1)/(Nu_{\infty}-1)$ in each group approximately follows a scaling $(Nu-1)/(Nu_{\infty}-1) \propto Ro^3$. It should be mentioned here that the derived scalings of $u''_{z}/{u''_{z}}$ and $(Nu-1)/(Nu_{\infty}-1)$ on $Ro$ are empirical results, which are currently lack of theoretical explanations. In the low Rossby regime, only a few data points are available for the fits. The robustness of these scalings needs to be checked when more data points are available in the future.

\subsection{Asymmetry between cyclones and anticyclones}
It has long been found that asymmetry between cyclones and anticyclones emerges in rapidly rotating convection \citep{chan2007rotating,kapyla2011starspots,guervilly2014large,guervilly2017jets}.
\citet{guervilly2014large} have discussed several possible mechanisms to explain the asymmetry. The most likely mechanism is that the thermal plumes ejected from the thermal boundary layers tend to induce more cyclonic vorticity. Therefore the clustering of the like-signed cyclonic vorticity favours the formation of the large-scale cyclone. In the low Rossby number limit, both ejection and injection of thermal plumes are allowed in the thermal boundary layer \citep{vorobieff2002turbulent,sprague2006numerical}. Thus it is expected that the asymmetry would disappear when $Ro$ is very small \citep{julien2012statistical}. \citet{guervilly2014large} have studied this mechanism for cases with dominant large-scale cyclone. Here we extend the discussion of this mechanism across different flow pattern regimes. Following the work of \citet{guervilly2014large}, we define the axial vorticity skewness as
\begin{eqnarray}
S(z)=\frac{ \int_{A} \overline{\omega_{z}^3}dA /\int_{A}dA }{\left(\int_{A} \overline{\omega_{z}^2}dA/\int_{A}dA \right)^{3/2}}~,
\end{eqnarray}
the $z$-dependent axial vorticity skewness as
\begin{eqnarray}
S'(z)=\frac{\int_{A} \overline{\left(\omega_{z} - \int_{z} \omega_{z} dz\right)^3} dA /\int_{A}dA  }{\left[\int_{A} \overline{\left(\omega_{z} - \int_{z} \omega_{z} dz\right)^2} dA/\int_{A}dA\right]^{3/2}}~,
\end{eqnarray}
and the $z$-invariant axial vorticity skewness as
\begin{eqnarray}
\overline{S}=\frac{\int_{A}\overline{(\int_{z} \omega_{z}dz)^3 }dA/\int_{A}dA}{\left[\int_{A}\overline{(\int_{z} \omega_{z}dz)^{2} }dA/\int_{A}dA\right]^{3/2}}~,
\end{eqnarray}
where $dA=dxdy$ is the differential area element in the horizontal planes.

\begin{figure}[!htbp]
\plotone{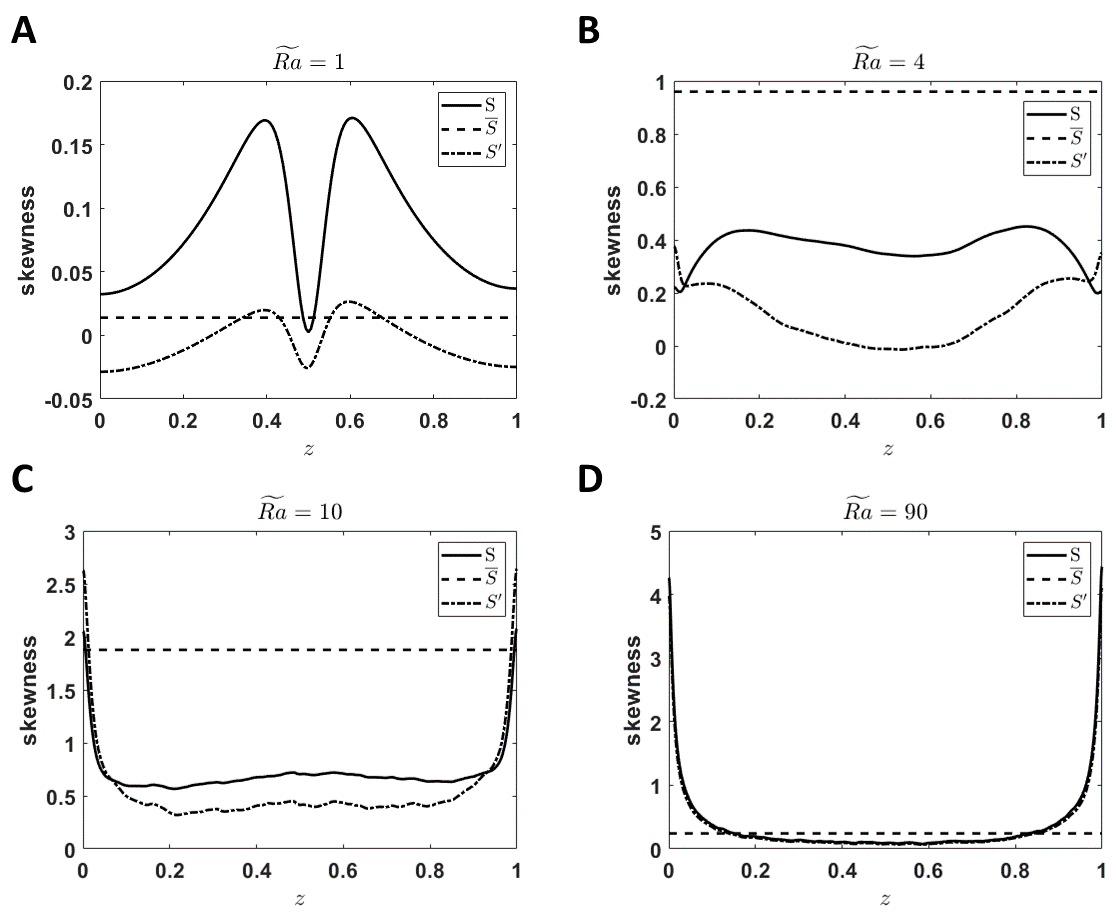}
\caption{The skewnesses $S$, $\overline{S}$, and $S'$ as functions of height. The four panels are for cases C1, C3, C5, and C13, respectively. \label{fig:fig11}}.
\end{figure}

Fig.~\ref{fig:fig11} shows $S(z)$, $S'(z)$, and $\overline{S}$ for cases C1, C3, C5, and C13. For case C13 (regime IV), $S(z)$ and $S'(z)$ have a similar structure (Fig.~\ref{fig:fig11}C), with a distribution of large positive values near the thermal boundary layers and small positive values (almost zeros) in the middle of the box. Fig.~\ref{fig:fig1} also shows that small but strong cyclonic vortex structures are created in the thermal boundary layers in this case. These vortex structures can only penetrate a short distance, thus the skewness profiles decrease rapidly away from the boundaries. For case C5 (regime III), a large-scale cyclone appears so that the $z$-invariant axial vorticity skewness $\overline{S}$ has a large positive value (Fig.~\ref{fig:fig11}C). $S(z)$ and $S'(z)$ have a similar profile near the boundaries. In the middle region, $S(z)$ are larger than $S'(z)$ but both of them are positive. From Fig.~\ref{fig:fig1}, we see that both cyclonic and anticyclonic vortex structures have been developed in the thermal boundary layers, but the cyclonic vortices are stronger and can penetrate farther away from the boundaries. For case C3 (regime II), although a pair of cyclone and anticyclone are formed, asymmetry between cyclone and anticyclone has not completely disappeared ($\overline{S}$ is about one in Fig.~\ref{fig:fig11}B). Similar to case C5, both cyclonic and anticyclonic vortex structures have been developed in the thermal boundary layers in case C3 (Fig.~\ref{fig:fig1}). However, the anticyclonic vortex structures in case C3 are stronger and can penetrate deeper than those in case C5. As a result, the $z$-dependent axial vorticity skewness $S'(z)$ is close to zero in the middle region (Fig.~\ref{fig:fig11}B). For case C1 (regime I), anticyclonic structures have similar strengths as cyclonic structures (Fig.~\ref{fig:fig1}). Since the vortical profile is almost antisymmetric about the middle plane (Fig.~\ref{fig:fig1}), $z$-invariant axial vorticity skewness $\overline{S}$ is nearly zero (Fig.~\ref{fig:fig11}A). $S(z)$ is positive in both lower and upper half boxes (Fig.~\ref{fig:fig11}A), which indicates that the vortical profile is not exactly antisymmetric (cyclonic vorticity is stronger than anticyclonic vorticity). $S'(z)$ turns to be negative near the boundaries in this case, but its value is too small to draw any affirmative conclusion.

In the study of \citet{vorobieff2002turbulent}, they showed that flow patterns are dominated by cyclonic thermal plumes when $Ro\sim 1$. When $Ro$ is further reduced, the number of anticyclonic thermal plumes increases but cyclonic thermal plumes are still favoured. Our simulation result agrees well with their experimental studies. Our result also supports the mechanism proposed by \citet{guervilly2014large}, that is, the preference for cyclonic thermal plumes might be responsible for the asymmetry between cyclones and anticyclones.

\subsection{Comparison with simulations at $Pr=1$}
\begin{figure}[!htbp]
\plotone{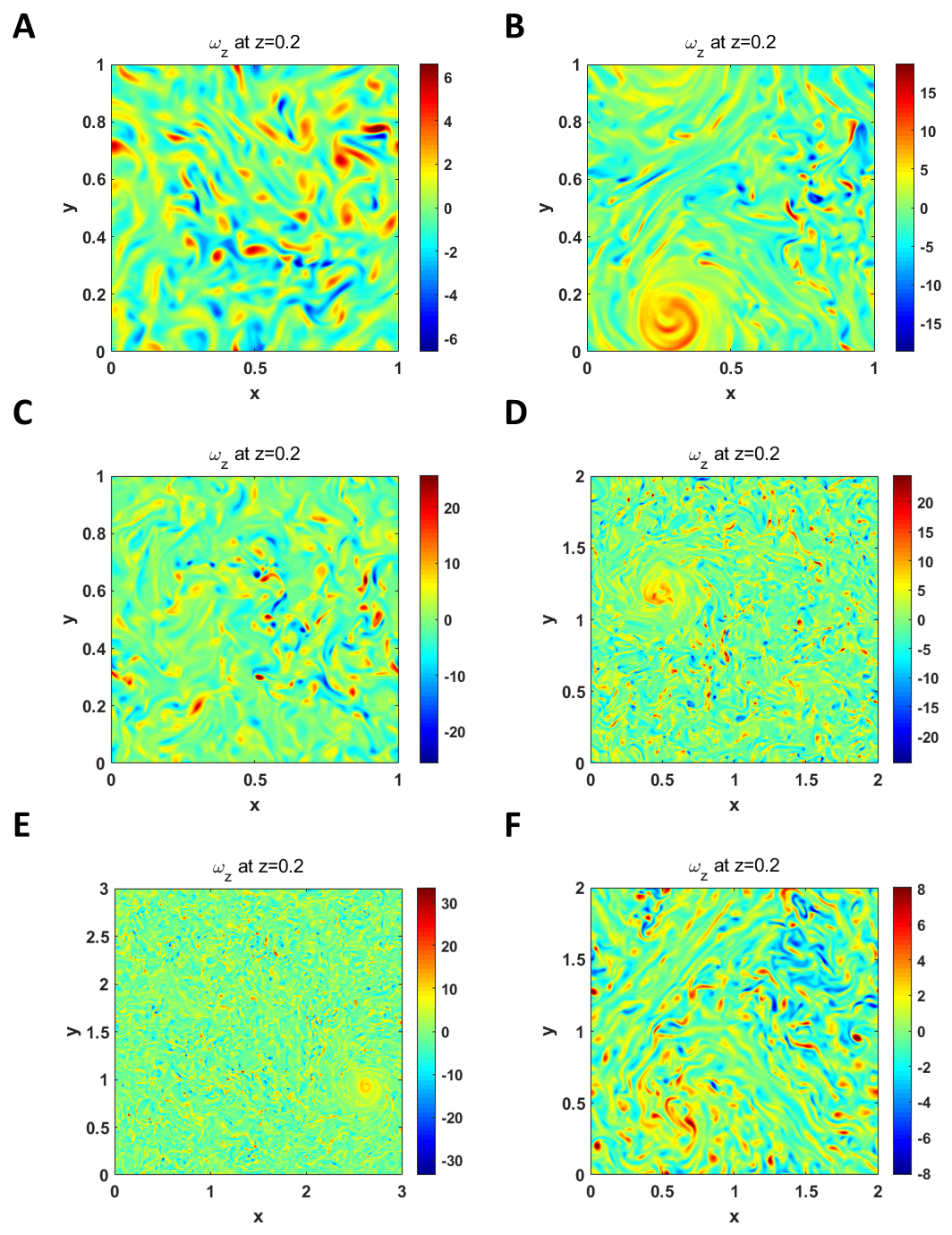}
\caption{Panels A-E show the horizontal cross-sections (at z=0.2) of the axial vorticity for cases D1, D3, D7, D7b, and D7c, respectively.}\label{fig:f12}
\end{figure}

From RBBC simulations at $Pr=1$, \citet{favier2014inverse} reported a higher critical convective Rossby number $Ro_{c1} \approx 0.6$ for the appearances of LSVs. In most of their simulations, the aspect ratios of simulation boxes are higher than one ($\Gamma>1$). \citet{guervilly2014large} has also reported that the aspect ratio has impact on the appearances of LSVs. To investigate the effects of $Pr$ and $\Gamma$, we have run several simulations at $Pr=1$ with different aspect ratios. The simulation parameters are listed in Table~\ref{table:tab2}. First, we notice that from cases D1-D7 that the criterions on the appearance of large-scale vortices at $Pr=0.1$ can also be applied to cases at $Pr=1$. That is, for $\Gamma=1$, the appearance of LSVs generally requires $Ro<0.4$ and $Re_{z}>400$. Fig.~\ref{fig:f12}A shows the flow structure of case D1. In this case, multiple small vorticies appear as the vertical Reynolds number is small. Fig.~\ref{fig:f12}B shows the flow structure of case D3. Since $Ro<0.4$ and $Re_{z}>400$, a large scale cyclone appears as expected. Fig.~\ref{fig:f12}C shows the flow structure of case D7. For this case, $Ro=0.55$ and there is no evidence for the appearance of large-scale vortices.

In the work of \citet{favier2014inverse}, they performed the same simulation as case D7, except that a different aspect ratio with $\Gamma=3$ was used. However, they have observed large-scale vortices in their simulation. Thus the aspect ratio probably is important on the formation of large-scale vortices \citep{favier2014inverse,guervilly2014large}. To investigate the effect of $\Gamma$, we have performed two companion simulations D7b and D7c with different aspect ratios $\Gamma=2$ and $\Gamma=3$, respectively. Figs.~\ref{fig:f12}D and \ref{fig:f12}E show the flow structures of D7b and D7c. Apparently, large-scale cyclones appear in these two cases. This is consistent with the result of \citet{favier2014inverse}. Therefore, aspect ratio indeed affects the appearance of large-scale vortices. Compared to simulations at $\Gamma=1$, the critical convective Rossby number on the appearance of LSVs is higher for simulations at larger $\Gamma$.

\citet{favier2014inverse} have also reported several cases with $Ro\leq 0.1$ at $Pr=1$ and $\Gamma=1$. In our simulations at $Pr=0.1$, we find that coexisted large-scale cyclone and anticyclone appear when $Ro<0.1$ and $Re_{z}$ is large enough. One may ask whether this criterion can be applied to simulations at higher $Pr$. To investigate this problem, we have run a simulation case at $Pr=1$ with $Ro=0.0548$ and $Ra=3\times 10^{9}$. This simulation case was also reported in \citet{favier2014inverse}, but here we have used a higher grid resolution. Fig.~\ref{fig:f12}F shows that coexisted large-scale cyclone and anticyclone indeed appear in this case. Thus the criterion obtained in low $Pr$ regimes is probably also valid in high $Pr$ regimes. Affirmative conclusion requires more simulations across different parameter regimes.

\begin{figure}[!htbp]
\plotone{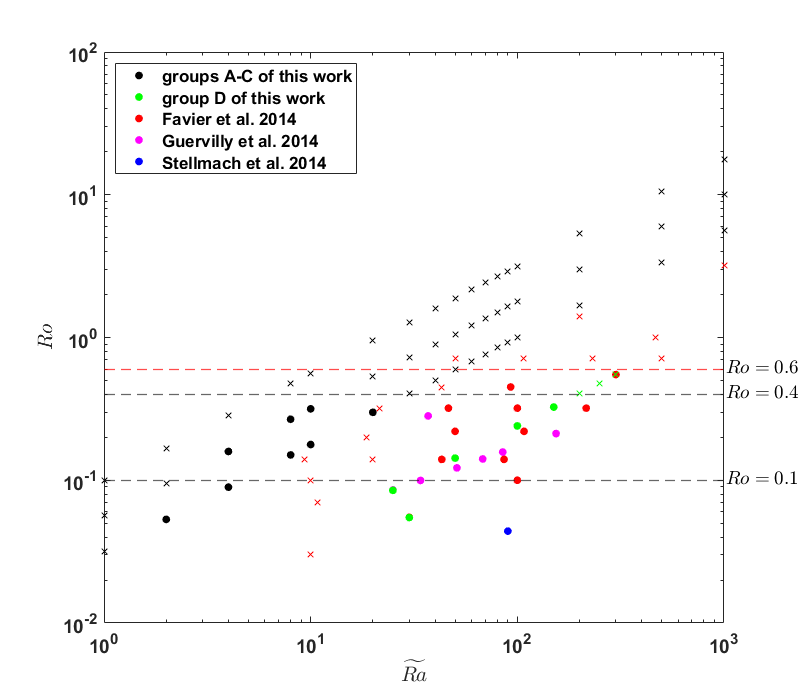}
\caption{Summary on simulation results from different studies. The black, green, red, megenta, and blue colors represent results obtained from groups A-C of this work, group D of this work, the work of \citet{favier2014inverse}, the work of \citet{guervilly2014large}, and the work of \citet{stellmach2014approaching}, respectively. The symbol cross denotes that no LSV appears, and the symbol dot denotes that LSVs appear.}\label{fig:f13}
\end{figure}

Fig.~\ref{fig:f13} presents a summary of simulation results from different studies. As shown in the figure, for the appearance of LSVs, it requires that $Ro$ is smaller than a certain critical value. We find this critical value is about $Ro\sim 0.4$ for a unit box, but \citet{favier2014inverse} reported a higher value of about $Ro\sim 0.6$ for a wider box. The variation of this critical $Ro$ on the size of box needs to be further investigated. A conjecture would be that the critical $Ro$ increases slightly with $\Gamma$. The appearance of coexisted cyclone and anticyclone occurs at a lower critical value $Ro\sim 0.1$ in a unit box. Interestingly, \citet{stellmach2014approaching} found a pair of cyclone and anticyclone at $Ro=0.044$ and $Pr=1$ in a small box with $\Gamma\approx 0.22$. If the critical $Ro$ indeed increases with $\Gamma$, then we would expect that the critical $Ro$ for the appearance of a pair of cyclone and anticyclone is slightly lower than 0.1 in $\Gamma=0.22$. The case in \citet{stellmach2014approaching} has a $Ro=0.044<0.1$, which still satisfies the criterion for the emergence of coexisted cyclone and anticyclone. From the simulation results, we speculate that the criterions for the appearances of large-scale vortices could possibly be universal for flows at different $Pr$.

\startlongtable
\begin{deluxetable*}{ccccccccccccccccccc}
\tablecaption{Parameters of simulation cases at $Pr=1$\label{table:tab2}}
\tablehead{
 Case & $\Gamma$ & $\ell_{c}$   & $u''$ & $u_{z}''$ & $u_{h}''$ & $Ra$ & $Re_{z}$ & $Ro$ & $E$ & $\widetilde{Ra}$ & $Nu$ & Regime
}
\startdata
D1  & 1 &  0.082 & 0.0606 & 0.0208 & 0.0567 & $3\times 10^8$  & 359.95    & 0.0850  & $4.905\times 10^{-6}$ & 25   & 7.463   & I    \\
D2  & 1 &  0.097 & 0.1510 & 0.0430 & 0.1445 & $3\times 10^8$  & 744.42    & 0.1429  & $8.249\times 10^{-6}$ & 50   & 21.259  & III  \\
D3  & 1 &  0.116 & 0.2228 & 0.0612 & 0.2140 & $3\times 10^8$  & 1060.08   & 0.2403  & $1.387\times 10^{-5}$ & 100  & 37.563  & III  \\
D4  & 1 &  0.128 & 0.2229 & 0.0690 & 0.2116 & $3\times 10^8$  & 1195.18   & 0.3257  & $1.880\times 10^{-5}$ & 150  & 46.023  & III  \\
D5  & 1 &  0.138 & 0.1497 & 0.0774 & 0.1272 & $3\times 10^8$  & 1339.81   & 0.4041  & $2.333\times 10^{-5}$ & 200  & 52.596  & IV   \\
D6  & 1 &  0.146 & 0.1422 & 0.0819 & 0.1148 & $3\times 10^8$  & 1419.17   & 0.4777  & $2.758\times 10^{-5}$ & 250  & 56.295  & IV   \\
D7  & 1 &  0.152 & 0.1369 & 0.0850 & 0.1054 & $3\times 10^8$  & 1472.57   & 0.5477  & $3.162\times 10^{-5}$ & 300  & 59.711  & IV   \\
D7b & 2 &  0.152 & 0.1882 & 0.0825 & 0.1685 & $3\times 10^8$  & 1428.41   & 0.5477  & $3.162\times 10^{-5}$ & 300  & 58.275  & III  \\
D7c & 3 &  0.152 & 0.2067 & 0.0818 & 0.1893 & $3\times 10^8$  & 1415.96   & 0.5477  & $3.162\times 10^{-5}$ & 300  & 57.983  & III  \\
D8  & 1 &  0.048 & 0.0724 & 0.0144 & 0.0709 & $3\times 10^9$  & 786.16    & 0.0548  & $1.000\times 10^{-6}$ & 30   & 9.246   & II
\enddata
\tablecomments{$\Gamma$ is the lateral-to-height aspect ratio; $\ell_{c}$ is the critical wavelength for the onset of convection; $u''$ is the rms velocity; $u_{z}''$ is the rms vertical velocity; $u_{h}''$ is the rms horizontal velocity; $Ra$ is the Rayleigh number; $Re_{z}$ is the vertical Reynolds number; $Ro$ is the convective Rossby number; $E$ is the Ekman number; $\widetilde{Ra}$ is the modified Rayleigh number; $Nu$ is the Nusselt number; and the cases are classified into four regimes according to flow patterns. Temporally and spatially (the whole box) averaged values are reported. The grid resolutions are $N_{x}\times N_{y}\times N_{z}=256\times 256 \times 257$ for D1-D7, $512\times 512 \times 257$ for D7b and D8, and $768\times 768 \times 257$ for D7c.}
\end{deluxetable*}

\section{Summary}
In this paper, we present results of numerical simulations on rapidly rotating Rayleigh-B\'enard convection at small Prandtl number $Pr=0.1$. We have demonstrated that LSVs (both cyclone and anticyclone) can be generated at moderate Rayleigh number with a small Prandtl number. Four different flow regimes are identified in our simulation: the regime of multiple small vortices, the regime of coexisted large-scale cyclone and anticyclone, the regime of large-scale cyclone, and the regime of turbulence. On the formation of LSVs, two conditions are required to be satisfied: high vertical Reynolds number and low Rossby number. Simulation result suggests that cyclone emerges when the convective Rossby number $Ro$ is smaller than 0.4; the appearance of coexisted large-scale cyclone and anticyclone requires $Ro$ to be smaller than 0.1. The Reynolds number is also crucial for the formation of LSVs. In the low Rossby range, the flow prefers to form as vertically aligned small vortices instead of LSVs if $Re_{z}<400$. The formation of LSVs requires energy transferred inversely from small scales to large scales. Based on the calculation of the transfer coefficient, we find that the kinetic energy can be directly transferred from small scales to largest scale. The analysis on 2D flow pattern shows that two separated cyclonic and anticyclonic regions coexist in regimes II and III. We have investigated the heat transfer efficiency within these two regions. The statistical results indicate that the heat transfer is more efficient in the anticyclonic region. The reason is that the effective rotation rate is smaller in the anticyclonic region than in the cyclonic region. Hence turbulence is more likely to be freely developed in the anticyclonic region. As a result, the turbulent heat transfer is more efficient in the anticyclonic region. We have also calculated the heat transfer by separating it into the 2D and 3D contributions. The result shows that the 2D contribution increases and 3D contribution decreases in transporting convective flux as rotation rate increases. The effects of rotation on statistical results of normalized velocities and modified Nusselt number (by the corresponding values of reference non-rotating cases) are discussed. We find that the normalized vertical velocity monotonically increases with increasing $Ro$. It indicates that rotation hinders vertical convection. In the regime of turbulence, the normalized vertical velocity appears to be irrelative to the Rayleigh number, but mainly dependent on $Ro$. The normalized vertical velocity tends to follow a 2/7 scaling with $Ro$ in the slowly rotating regime, and a 5/3 scaling with $Ro$ in the rapidly rotating regime. For the heat transfer, we find that the Nusselt number can be drastically reduced with increasing rotation when large-scale vortices or multiple small vortices appear.

Our results may have interesting implications to rapidly rotating gas giants, such as Jupiter and Saturn. In these planets, the Prandtl numbers are estimated to be about 0.1, the convective Rossby numbers of these planets are much smaller than 0.1, and the vertical Reynolds numbers are much higher than 400 \citep{schubert2011planetary}. Thus, it is anticipated that convectively driven large-scale cyclones and anticyclones can be generated in these rapidly rotating planets. The large-scale vortices observed in these gas giants are probably driven by rapidly rotating convection sustained by the internal heat. It has to be admitted that there is still a gap between the parameters we used in this paper and those of real Jupiter and Saturn. Although a relative simple Boussinesq model is used, the results provide helpful insights on understanding the dynamics of these gas giants. The model we used in this paper can be improved in several aspects. First, here we only consider an incompressible flow, while the atmospheres of these planets are compressible. It seems that compressibility has important effect on the appearance and size of large-scale vortices \citep{kapyla2011starspots,chan2013numerical}. The effect of compressibility is worth to be investigated in the future. Second, the Rayleigh numbers we simulated are too low compared to the realistic value of these planets. There is no guarantee that the criterions on the appearance of large-scale vortices obtained in low Rayleigh region can be extended to high Rayleigh regions. It is worthwhile to perform simulations with higher Rayleigh numbers in the future. Third, in this paper, we focus our discussion on an lateral to height aspect ratio of one. It has been found that the aspect ratio has important effects on the formation on large-scale vortices \citep{guervilly2014large}. Our preliminary study on the effect of aspect ratio indicates that the critical convective Rossby number for the occurrence of large-scale vortices might increase with the aspect ratio. Due to the computational difficulty, most simulations are performed in small or moderate size boxes. In large boxes, multiple large-scale vortices may appear. For example, recent simulations \citep{cai2021deep} in a high aspect ratio (lateral to height aspect ratio is 16) box have successfully produced polygonal structures of large-scale vortices. Detailed discussion on the effects of aspect ratio is very expensive. We decide to postpone the exploration to future researches.

\begin{acknowledgments}
I thank the reviewer for helpful suggestions on improving this manuscript. I conceived the idea when I visited the UCLA SpinLab three years ago, and I thank Jonathan Aurnou and the lab members for interesting discussions on rotating Rayleigh-B\'enard convection. I appreciate Jing Li and David Jewitt for their hospitality during my visit. I am also grateful to Kwing L. Chan for fruitful discussions on rotating convection in the past years. This work was partially supported by the Guangdong Basic and Applied Basic Research Foundation (No.2019A1515011625), NSFC (Nos.12173105, 11503097), Science and Technology Development Fund, Macau SAR (Nos.0045/2018/AFJ, 0156/2019/A3), the China Space Agency Project (No.D020303), and the China Manned Space Project (No.CMS-CSST-2021-B09). The simulations were performed in the supercomputers at the Purple Mountain Observatory, the National Supercomputer Center in Guangzhou, and the Macau University of Science and Technology.
\end{acknowledgments}

\appendix\section{Supplementary Movies}
\begin{figure}[!htbp]
\begin{interactive}{animation}{regimeI.mpg}
\plotone{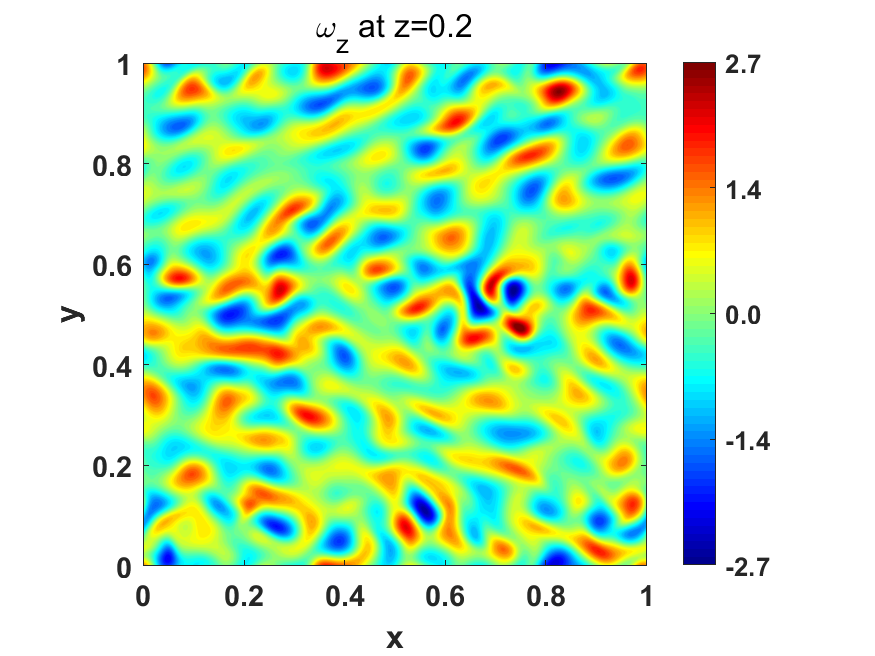}
\end{interactive}
\caption{The animation shows the time evolution of the vertical vorticity structure at $z=0.2$ for the case C1 (regime I). The time period for the movie is about 37.7 units of time.}\label{fig:regimeI}
\end{figure}

\begin{figure}[!htbp]
\begin{interactive}{animation}{regimeII.mpg}
\plotone{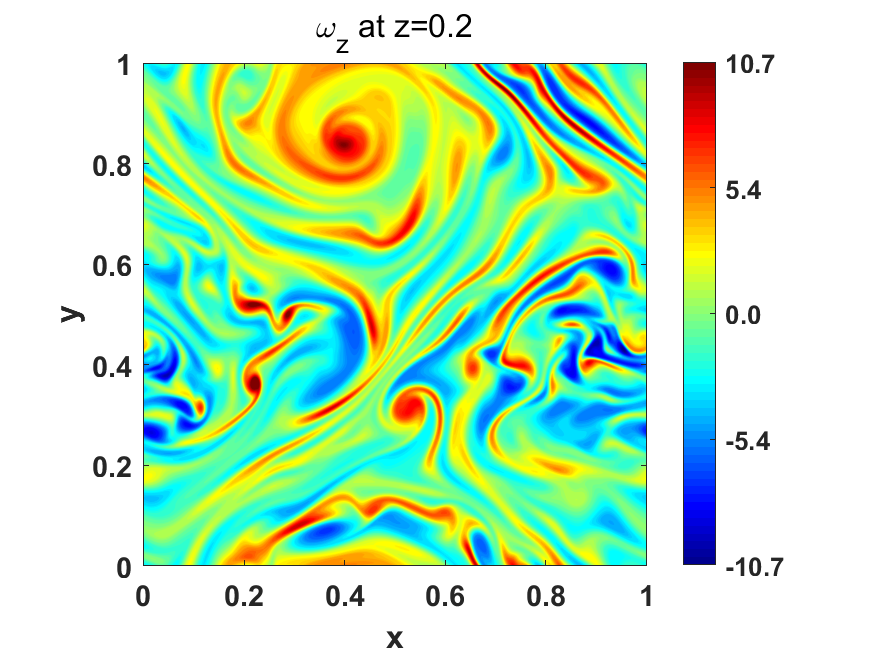}
\end{interactive}
\caption{The animation shows the time evolution of the vertical vorticity structure at $z=0.2$ for case C3 (regime II). The time period for the movie is about 37.7 units of time.}\label{fig:regimeII}
\end{figure}

\begin{figure}[!htbp]
\begin{interactive}{animation}{regimeIII.mpg}
\plotone{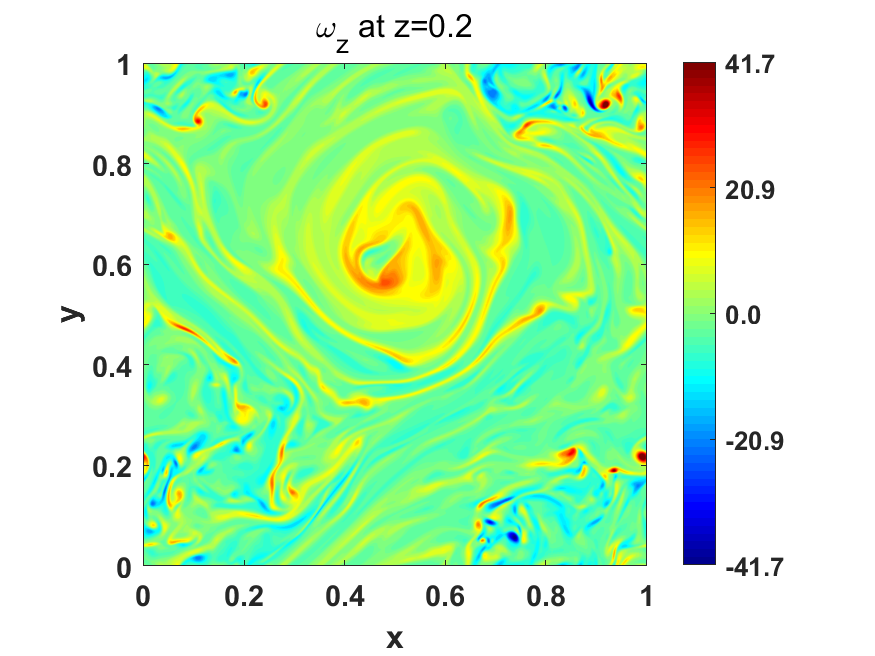}
\end{interactive}
\caption{The animation shows the time evolution of the vertical vorticity structure at $z=0.2$ for case C5 (regime III). The time period for the movie is about 37.7 units of time.}\label{fig:regimeIII}
\end{figure}

\begin{figure}[!htbp]
\begin{interactive}{animation}{regimeIV.mpg}
\plotone{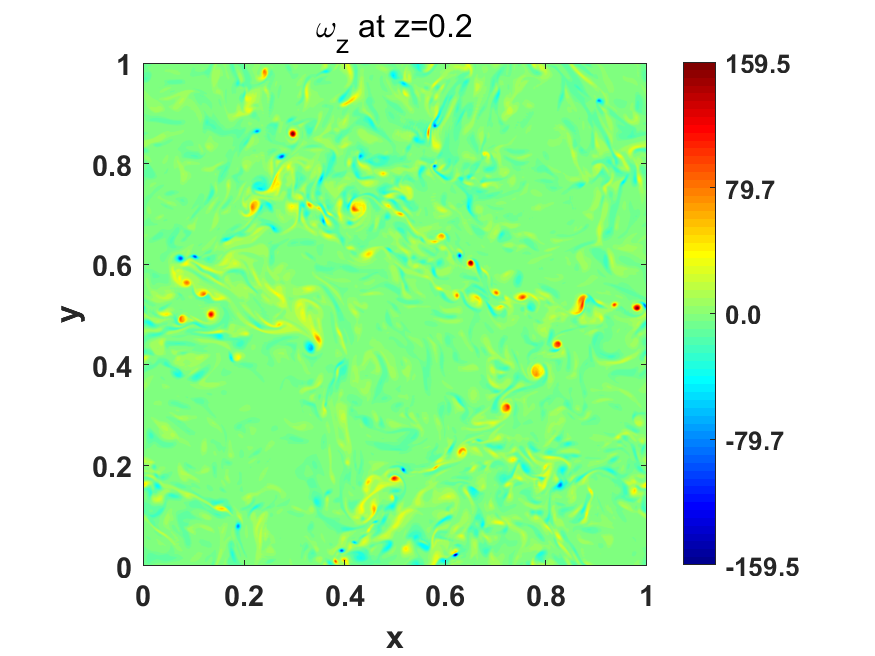}
\end{interactive}
\caption{The animation shows the time evolution of the vertical vorticity structure at $z=0.2$ for case C13 (regime IV). The time period for the movie is about 37.7 units of time.}\label{fig:regimeIV}
\end{figure}

\begin{figure}[!htbp]
\begin{interactive}{animation}{rot_regimeI.mpg}
\plotone{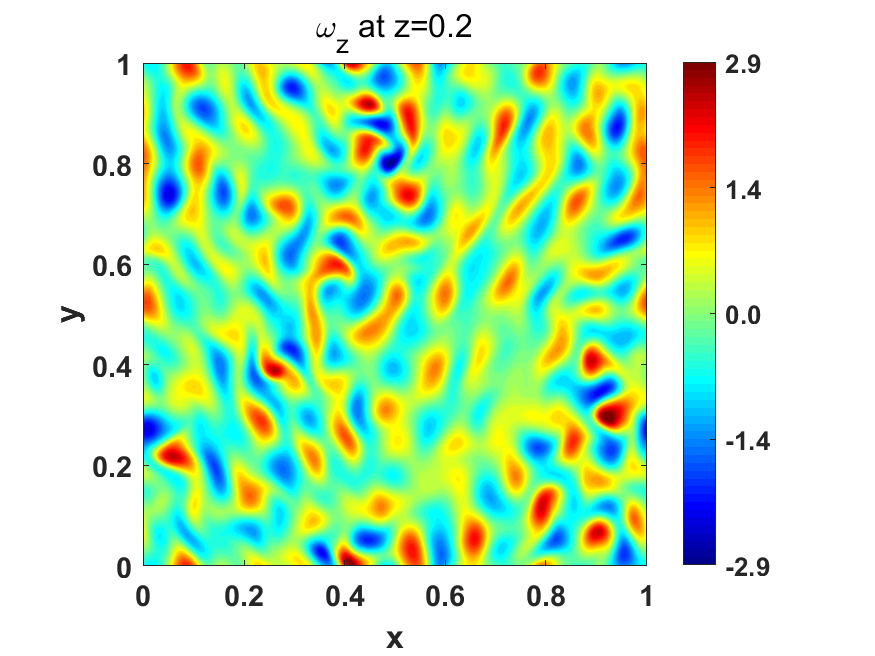}
\end{interactive}
\caption{A companion simulation of case C1 by rotating the domain by $90^{\circ}$. The time period for the movie is about 37.7 units of time.}\label{fig:rot_regimeI}
\end{figure}

The animations for Figs.~\ref{fig:regimeI}-\ref{fig:regimeIV} show the time evolutions of vertical vorticity structures at $z=0.2$ for cases C1, C3, C5, and C13, respectively. The time period for each movie is about 37.7 units of time. The animation for Fig.~\ref{fig:rot_regimeI} shows a companion simulation of case C1 by rotating the domain by $90^{\circ}$.

\bibliographystyle{aasjournal}
\bibliography{rrc}



\end{document}